\documentclass[11pt]{article}
\usepackage{jheppub}

\usepackage{mathtools}

\usepackage{indentfirst}

\usepackage{hyperref}
\hypersetup{
	unicode,	
	colorlinks,
	citecolor=[rgb]{0,.3,.5}, 
	linkcolor=[rgb]{0,.3,.5},
	urlcolor=[rgb]{0,.3,.5}, 
	bookmarksopen=true,
	bookmarksopenlevel=\maxdimen,
}

\usepackage{amsmath, amssymb} 
\usepackage{mathrsfs} 			
\usepackage{bm}				

\def\ul#1{\underline #1}
\def\vev#1{{\langle #1\rangle}}

\def\on#1#2{{\buildrel{\mkern2.5mu#1\mkern-2.5mu}\over{#2}}}

\def\dt#1{\smash{\on{\hbox{\bf .}}{#1}}\vphantom{#1}}     

\mathcode`\*="702A                  
\def\half{{\textstyle{1\over{\raise.1ex\hbox{$\scriptstyle{2}$}}}}}

\newcommand{\cA}{\mathcal{A}}
\newcommand{\cD}{\mathcal{D}}
\newcommand{\cE}{\mathcal{E}}
\newcommand{\cF}{\mathcal{F}}
\newcommand{\cG}{\mathcal{G}}

\newcommand{\cL}{\mathcal{L}}
\newcommand{\cO}{\mathcal{O}}
\newcommand{\cU}{\mathcal{U}}
\newcommand{\cV}{\mathcal{V}}
\newcommand{\cW}{\mathcal{W}}

\newcommand{\HC}{\text{c.c.}}
\newcommand{\pa}{\partial}

\newcommand{\dalpha}{{\dot\alpha}}
\newcommand{\dbeta}{{\dot\beta}}

\newcommand{\ha}{{\hat a}}
\newcommand{\hb}{{\hat b}}
\newcommand{\hc}{{\hat c}}
\newcommand{\hd}{{\hat d}}
\newcommand{\he}{{\hat e}}

\newcommand{\hA}{{\hat A}}
\newcommand{\hB}{{\hat B}}
\newcommand{\hC}{{\hat C}}
\newcommand{\hD}{{\hat D}}

\newcommand{\hH}{{\hat H}}

\newcommand{\hM}{{\hat M}}
\newcommand{\hN}{{\hat N}}
\newcommand{\hP}{{\hat P}}

\newcommand{\halpha}{{\hat\alpha}}
\newcommand{\hbeta}{{\hat\beta}}
\newcommand{\hgamma}{{\hat\gamma}}

\newcommand{\eps}{\epsilon}

\newcommand{\eol}{\nonumber \\}

\numberwithin{equation}{section}

\title{
$\bm{N=1}$ 
Supercurrents of Eleven-dimensional Supergravity}

\author{Katrin Becker,}
\author{Melanie Becker,}
\author{Daniel Butter,}
\author{and William D. Linch III}
\affiliation{
George P. and Cynthia Woods
Mitchell Institute for 
Fundamental Physics and Astronomy, \\
Texas A\&{}M University.\\
College Station, TX 77843, USA}

\emailAdd{kbecker@physics.tamu.edu}
\emailAdd{mbecker@physics.tamu.edu}
\emailAdd{dbutter@tamu.edu}
\emailAdd{wdlinch3@gmail.com}

\preprint{MI-TH-1877}

\arxivnumber{1803.00050}

\abstract{Eleven-dimensional supergravity can be formulated in superspaces locally of the form $\bm X\times Y$ where $\bm X$ is 4D $N=1$ conformal superspace and $Y$ is an arbitrary 7-manifold admitting a $G_2$-structure.
The eleven-dimensional 3-form and the stable 3-form on $Y$ define the lowest component of a gauge superfield on $\bm X \times Y$ that is chiral as a superfield on $\bm X$.
This chiral field is part of a tensor hierarchy giving rise to a superspace Chern-Simons action and its real field strength defines a lifting of the Hitchin functional on $Y$ to the $G_2$ superspace $\bm X\times Y$. These terms are those of lowest order in a superspace Noether expansion in seven $N=1$ conformal gravitino superfields $\Psi$. 
In this paper, we compute the $O(\Psi)$ action to all orders in the remaining fields.
The eleven-dimensional origin of the resulting non-linear structures is parameterized by the choice of a complex spinor on $Y$ encoding the off-shell 4D $N=1$ subalgebra of the eleven-dimensional super-Poincar\'e algebra.
}

\begin{document}
\maketitle

\section{Introduction}
\label{S:Intro}

The component formulation of eleven-dimensional supergravity \cite{Cremmer:1978km} belies a startling amount of structure present in this enigmatic theory.
Most famously, this theory is invariant under the maximal number of local supersymmetries. 
The maximal supersymmetry can be made manifest in superspace \cite{Cremmer:1980ru, Brink:1980az}.
When the amount of Poincar\'e symmetry manifest is not required to be maximal, additional/alternative symmetries can be realized linearly---most famously, the local $SU(8)$ \cite{deWit:1986mz} and 
global exceptional 
symmetries \cite{Hohm:2013pua, Hohm:2013uia, Godazgar:2014nqa}.
These can also be realized in superspace \cite{Brink:1979nt} when the theory is truncated to four dimensions \cite{Cremmer:1978ds, Cremmer:1979up}.
As with all theories representing more than nine supercharges, the super-Poincar\'e algebras close only up to the component field equations of motion \cite{Berkovits:1993hx}.
That is, these supersymmetries and the superspaces in which they are manifest are ``on-shell''. 

An alternative when only a part of the Poincar\'e algebra is required to be manifest is to represent only the supersymmetries associated to the subalgebra linearly. 
That is, we may contemplate describing the eleven-dimensional supergravity theory as a fibration by simpler superspaces.
The simplest such choice (or at least the most well-developed by far) is 4D $N=1$ superspace.
This superspace has many user-friendly features (e.g. off-shell, finitely many auxiliary fields, chiral representations) not the least of which is being the most relevant phenomenologically. 
We therefore propose to study eleven-dimensional supergravity in superspaces that are locally of the form $\bm X \times Y$ where $Y$ is any Riemannian 7-manifold and $\bm X$ is a curved 4D $N=1$ superspace.
This program was initiated in \cite{Becker:2016xgv, Becker:2016rku, Becker:2016edk, Becker:2017njd, Becker:2017zwe}.

In this paper, we extend the construction of eleven-dimensional supergravity in superspace by deriving the supercurrents for the 4D $N=1$ conformal supergravity and seven conformal gravitino prepotentials (the spin-2 and spin-$\frac32$ parts) to all orders in the remaining (spin $\leq1$) superfields.
This is a necessary step in the determination of the action to all orders in the gravitino superfield. 
Additionally, it elucidates the eleven-dimensional origin of the conformal compensator fields: 
When supergravity is na\"ively switched off, these are superfluous component fields required for the superspace embedding of the 3-form hierarchy. 
However in the superspace splitting $\bm X \times Y$, local superconformal symmetries emerge in addition to those inherited from eleven dimensions. 
The consistency of our superfield description hinges on the fact that precisely the seemingly-superfluous components ``compensate'' for these fake symmetries. (They are their St\"uckelberg fields.)
We stress that the conformal symmetry is in the $\bm X$ factor and arises as an artifact of the splitting.\footnote{Attempts have been made to define the superconformal symmetry directly in eleven-dimensional superspace \cite{Howe:1997he, Gates:2001zz}, and to use it to represent all 32 supersymmetries off shell
\cite{Cederwall:2000ye, Gates:2001hf}.
} 

In section \ref{S:Review}, we review the results of our previous work \cite{Becker:2016xgv, Becker:2016rku, Becker:2016edk, Becker:2017njd, Becker:2017zwe} on the embedding of eleven-dimensional supergravity into off-shell superspaces $\bm X \times Y$ locally of the form $\mathbf R^{4|4} \times \mathbf R^7$. 
The component fields of eleven-dimensional supergravity consist of an elf-bein, a gravitino, and a 3-form gauge field. 
Decomposing these fields in a $(4+7)$-dimensional split leads to the embedding of the 3-form components into a collection of 4D $N=1$ $p$-form superfields with $p=0,\dots,3$ that are $q$-forms with $q=3-p$ in the additional seven directions. The original eleven-dimensional abelian gauge symmetry with 2-form gauge parameter decomposes into a set of gauge transformations in 4+7 dimensions that transform these superfields into each other. In addition, the $p$-forms in the tensor hierarchy are charged under the Kaluza-Klein vector gauging diffeomorphisms along $Y$. This gives rise to a non-abelian gauging of the tensor hierarchy and a Chern-Simons-like action which we embedded in superspace \cite{Becker:2016xgv, Becker:2016rku, Becker:2017njd}.

One of the component fields in the tensor hierarchy is a $[0,3]$-form, that is a  0-form in spacetime $X$ and a 3-form in the internal space $Y$. It is naturally contained in the bottom component of a chiral multiplet; since this bottom component is required by supersymmetry to be complex, the pseudo-scalar $[0,3]$-form is naturally paired with a scalar $[0,3]$-form which is invariant under the abelian gauge symmetry. This object defines a symmetric bilinear form that can be taken to be positive-definite for generic 7-manifolds $Y$ \cite{Hitchin2001}. As such, it defines a Riemannian volume on $Y$, and the superfield containing it defines a K\"ahler potential on the superspace. Modified by a certain function of another of the tensor hierarchy superfields, this defines a second $N=1$ supersymmetric action which, together with the Chern-Simons action, describes the dynamics of fields with 4D spin $\leq 1$ \cite{Becker:2016edk}.

A peculiar feature of this action is its invariance under global 4D $N=1$ superconformal transformations, which include both scale (Weyl) and chiral $U(1)_R$ transformations. It was argued in \cite{Becker:2016edk} that this is naturally enhanced to a local superconformal symmetry when coupling to $N=1$ supergravity.
It was shown in \cite{Becker:2017zwe} that the coupling to $N=1$ supergravity and the additional seven gravitino multiplets could be achieved at the linearized level -- that is, considering the action quadratic in fluctuations about an on-shell 4D Minkowski background with an arbitrary internal manifold of $G_2$ holonomy and vanishing 4-form flux. The explicit $N=1$ superspace action was given and its component action was shown to match that obtained directly from 11D. 

Our goal in this paper is to proceed beyond the approximations of \cite{Becker:2016edk, Becker:2017zwe} by constructing the coupling of both $N=1$ conformal supergravity and additional seven gravitino superfields to the non-linear action of the fields of lower 4D spin.
Specifically, we will expand in superfields (denoted $H^a$ and $\Psi^\alpha_i$ below) containing component spins $\geq \frac32$ but keeping the exact non-linear dependence on the remaining superfields at each order. 
To lowest order, this was done in reference \cite{Becker:2016edk} and checked to reproduce the non-linear scalar potential upon component projection. 
In section \ref{S:GravitinoSupercurrent}, we extend this result to the leading order in the seven gravitino superfields $\Psi^\alpha_i$ by constructing the complete supercurrent $J_\alpha^i$ (cf.\ \ref{E:DefJ.v1}), and motivate its derivation with a careful analysis of the symmetries of the conformal gravitino superfields. 
The intricate compensator mechanism associated to this superconformal symmetry has a strikingly simple interpretation when it is derived from eleven dimensions. As we show in section \ref{S:11DConnection}, reducing the 11D $N=1$ superspace frame to 4D $N=1$ can be done in different ways parameterized by a complex scalar and a complex spinor of $SO(7)$. By studying the gauge transformations of this parameterization, we discover that the bilinears of the spinors (dressed appropriately with the modulus and phase of the scalar) reproduce the transformation rules defining the superconformal compensators. 

With the gravitino couplings thus understood, we turn in section \ref{E:CSGcoupling} to their  superconformal graviton analog. 
Similarly to the superconformal gravitino multiplet, our analysis has been perturbative in the $Y$-dependence of the 4D $N=1$ superconformal graviton:
In references \cite{Becker:2016edk, Becker:2017njd} the 4D $N=1$ SG part was treated non-linearly as a $Y$-independent background. In reference \cite{Becker:2017zwe}, $Y$-dependent fluctuations were studied but only to quadratic order in all fields.
In section \ref{E:CSGcoupling}, we begin to address this point by giving the transformations of the gauged tensor hierarchy fields under the 4D $N=1$ superconformal symmetry to the first non-trivial order, now with $Y$-dependent gauge parameter.
We then derive the complete 4D $N=1$ conformal supercurrent by requiring invariance under these extended symmetries.

This completes the superspace construction to lowest non-trivial order in the spin-2 and $\tfrac32$ components including dependence on all eleven dimensions but treating the spin $\leq 1$ components exactly. 
To go beyond this order in spin $\geq \frac32$ fields requires understanding the non-linear terms in the gravitino expansion. In section \ref{S:CovariantForm} we take the first step in this direction by constructing modifications of the hierarchy field strengths that are invariant under part of the seven extended superconformal symmetries.

\section{Review and Summary}
\label{S:Review}

We begin by reviewing our previous results \cite{Becker:2016xgv, Becker:2016rku, Becker:2016edk, Becker:2017njd, Becker:2017zwe} on the description of eleven-dimensional supergravity in terms of superspaces of the form $\bm X \times Y$ where $\bm X$ is a 4D $N=1$ supermanifold and $Y$ is a real 7-manifold. 
Locally $\bm X$ is of the form $\mathbf R^{4|4}$ with coordinates $(x^m, \theta^\mu, \bar \theta{}^{\dt \mu})$ and indices $m=0,\dots, 3$ and $\mu, \dt \mu =1,2$ from the middle of the alphabets. Following the early/late convention, tangent indices are taken from the beginnings (e.g. $a$, $\alpha$, and $\dt \alpha$). 
Local coordinates on $Y$ will be denoted by $y^i$ with $i=1,\dots, 7$; we will generally not need tangent indices for $Y$. 
The body of $\bm X$ (i.e. its bosonic part) will be denoted by $X$ and eleven-dimensional indices on $X\times Y$ will be denoted in bold so that, for example, $x^{\bm m} = (x^m, y^i)$.

In succession we discuss the embedding of the components of the metric (\S{}\ref{S:Review:Metric}) and 3-form (\S{}\ref{S:Review:3form}) and their superspace gauge transformations, field strengths and Bianchi identities. 
With these ingredients, we build an invariant action consisting of a superspace volume term and a Chern-Simons-like term (\S{}\ref{S:Review:Action}).
We conclude our review by introducing the gravitino superfields and defining the gravitino expansion (\S{}\ref{S:Review:Grino}). 

\subsection{Decomposition of the metric}
\label{S:Review:Metric}

Let's first discuss the fields that arise from decomposing the 11D metric, as these will play a role in defining the covariant derivatives in 4D $N=1$ superspace. We employ the standard Kaluza-Klein decomposition
\begin{align}
g_{\bm m \bm n} =
\begin{pmatrix}
g_{m n} + \cA_m{}^i \cA_n{}^j g_{i j} & \cA_{m}{}^j g_{j i} \\
g_{i j} \cA_n{}^j & g_{i j}
\end{pmatrix}
.
\end{align}
A priori, the 4D metric $g_{mn}$, the Kaluza-Klein gauge field $\cA_m{}^i$, and the $Y$ polarizations of the metric $g_{ij}$ each depend on all eleven coordinates.

The 4D metric $g_{mn}$ is encoded in a real superfield
$H_{\alpha \dalpha} = (\sigma^a)_{\alpha \dalpha}H_{a} $ with a linearized gauge transformation
\begin{align}
\label{E:deltaHa}
\delta H_{\alpha \dalpha}  = \bar D_{\dalpha} L_\alpha - D_\alpha \bar L_{\dalpha} ~.
\end{align}
This defines it as an irreducible superspin-$\tfrac32$ representation: At the component level it contains the spin-2 polarizations of the frame $e_m{}^a$, the $N=1$ gravitino $\psi_{m}{}^{\alpha}$, and an auxiliary vector field $d_m$. The gauge symmetry \eqref{E:deltaHa} actually encodes local $N=1$ superconformal transformations, so that the frame is defined up to an overall Weyl rescaling, the gravitino is defined up to a shift in its spin-1/2 part corresponding to the action of the special superconformal $S$-supersymmetry, and $d_m$ is defined up to chiral $U(1)_R$ transformations for which $d_m$ is the gauge field.

In \cite{Becker:2017zwe}, the superfield $H_a$ appeared explicitly, but this quickly becomes unwieldy when going beyond quadratic order. It is usually much simpler to employ a curved Wess-Zumino-type superspace where the prepotential $H_a$ is encoded in the super-vielbein $E_M{}^A$. The natural superspace to employ is 4D $N=1$ conformal superspace \cite{Butter:2009cp}, where the $N=1$ superconformal symmetry is explicitly gauged and which describes precisely (in super-geometric language) the field content of $H_a$. However, there is a caveat to employing this superspace:  \emph{We have to assume that the super-vielbein (and its prepotential $H_a$) do not depend on $y$, since in the language of \cite{Butter:2009cp}, the super-vielbein is gauge invariant under any internal gauge symmetries.}
In the context of our split spacetime, these include the $GL(7)$ diffeomorphisms of $Y$. 
There is no technical obstruction to developing a superspace that relaxes this condition, but it does not yet exist, and we will revisit this point in section \ref{S:CovariantForm}.\footnote{Actually, this is not as constraining a scenario as it might seem. Because the $N=1$ supergeometry describes the metric only up to Weyl rescalings, if the external metric $g_{mn}(x,y)$ factorizes as $g_{mn} \rightarrow e^{\varphi(x,y)} g_{mn}(x)$, with $y$-dependence sequestered in a conformal factor, then the super-vielbein can be taken to just describe the $y$-independent piece. In realistic scenarios, this would describe the actual background configurations of interest. Then an explicit $y$-dependent $H_a$ could be introduced to describe general $y$-dependent \emph{fluctuations} about that background.
}

The non-abelian Kaluza-Klein gauge fields $\cA_m{}^i$ are described by a real unconstrained prepotential $\mathcal V^i$. Just as with the supergravity prepotential, it is more convenient to use a covariant description where the Kaluza-Klein multiplet is encoded in super-connections $\cA_M{}^i$ which covariantize the $N=1$ superspace derivatives $\nabla$, so that the super-connections transform as $\delta \cA^i = \nabla \tau^i$ with $\tau^i$ a real superfield describing internal diffeomorphisms. The Kaluza-Klein connections deform the algebra of the conformal superspace derivatives by introducing the curvature terms
\begin{align}
[\nabla_a, \bar \nabla_{\dalpha}] = - (\sigma_a)_{\alpha \dalpha} \mathscr L_{\mathcal W^{\alpha}} + \cdots
\end{align} 
where $\cW_\alpha{}^i$ is a chiral field strength obeying the standard conditions
$\bar\nabla_\dalpha \cW_\alpha{}^i = 0$ and 
$\nabla^\alpha \cW_\alpha{}^i = \bar\nabla_\dalpha \cW^\dalpha{}^i$.
Here $\mathscr L_v$ denotes the Lie derivative along the vector field $v\in \mathfrak X(Y)$. On forms it decomposes as $\mathscr L_v  = \partial \iota_v + \iota_v \partial$ into the de Rham differential $\partial$ on $Y$ and the interior product (contraction) with the vector field $v$. In contrast to the super-vielbein, the connection $\cA$ may depend on $y$.

The remaining $Y$ polarizations $g_{ij}$ of the metric are 28 real scalars from the point of view of $X$. 
We will address their embedding into superfields presently.

\subsection{Decomposition of the 3-form}
\label{S:Review:3form}

The eleven-dimensional 3-form splits up into a ``tensor hierarchy'' of $p$-forms,
\begin{align}
C_{\bm{mnp}} \rightarrow 
 C_{m n p}
 ~, C_{m n \,k}
~, C_{m\, j k}
~, C_{ijk}
\end{align}
which are embedded into a tower of $p$-form superfields. (The remaining 28 scalars from $g_{ij}$ will also be embedded in these fields.) 
Being forms also in the seven directions, they are charged under the non-abelian gauge field. The abelian part of the gauge transformation is parameterized by the superfields $\Lambda_{ij}$ (chiral), $U_i$ (real), and $\Upsilon_\alpha$ (chiral) encoding the components of an eleven-dimensional super-2-form.
Their weights are summarized in table \ref{T:Superscale}.
The non-abelian part $\mathfrak g = \mathfrak {diff}(Y)$ acts by the Lie derivative with respect to the real scalar superfield $\tau^i$.

The superfields of the hierarchy (not including the KK vector just described) are as follows (we mention only the embedding of bosons, summarized in table \ref{T:PotentialSpectrum}):
\begin{table}[t]
{\renewcommand{\arraystretch}{1.2} 
\begin{center}
\hspace{-5mm}
\begin{tabular}{|c|c|c|c|c|c|c|}
\hline
	& 3-forms 	& 2-forms & vectors		& scalars 	& auxiliaries \\
\hline
$X$	& $C_{mnp}$	& | 		& | 			& $G$				& $d_X$ \\
$\Sigma^\alpha_i$ & |& $C_{mn\,i}$& |		 	& $H_{i}$			& | \\
$V_{i j}$ & |		& |		& $C_{m \,i j}$	& |					& $d_{i j}$ \\
$\Phi_{ijk}$& |	& |		& |			& $C_{i j k}$, $F_{i j k}$ & $f_{i j k}$\\
$ \cV^i$ & |		& |		& $g_{m i}$	& |					& $\bm{d}^i$ \\
\hline
\end{tabular}
\end{center}
} 
\begin{caption}{Bosonic field content of the Chern-Simons prepotentials}
\label{T:PotentialSpectrum}
\footnotesize
We list the bosonic field content of of the gauged Chern-Simons superfield hierarchy. The bosons $G$, $H_i$, and seven of the $F_{ijk}$ can all be removed by a choice of Wess-Zumino gauge (cf.\ \S{}\ref{S:Review:Grino}).
\end{caption}
\end{table}
\begin{description}
	\item[Scalars] There are 35 chiral fields $\Phi_{ijk}$ containing the 35 pseudo-scalars $C_{ijk}$ from the 3-form and 28 metric scalars $g_{ij}$. (The remaining seven scalars will be shown to be pure gauge.)
These transform under non-abelian internal diffeomorphisms and under the abelian tensor hierarchy gauge transformations as
\begin{align}
\delta \Phi &= 
	\mathscr L_\tau \Phi 
	+ \partial \Lambda ~.
\end{align}	
It has two field strengths $E_{ijkl}$ (chiral) and $F_{ijk}$ (real)
\begin{align}
E = \partial \Phi
~~~\textrm{and}~~~
F = \tfrac1{2i}\left( \Phi - \bar \Phi\right) - \partial V~.
\end{align}
($V_{ij}$ is the vector multiplet prepotential we introduce next.)
The lowest component of $F_{ijk}$ is related to a Riemannian metric $g_{ij}$ by
\begin{align}
\label{E:metric}
\sqrt{g} \,g_{ij} :=  -\tfrac1{144} \epsilon^{klmnpqr} F_{ikl} F_{mnp} F_{jqr}~.
\end{align}
As we explain in detail in section \ref{S:11DConnection}, this metric is a generalization of the $G_2$ structure metric and $F_{ijk}$ that of a $G_2$ structure for the internal manifold $Y$.
It is not in general closed or co-closed.
	\item[Vectors] There are 21 real, unconstrained vector superfields $V_{ij}$ containing vector fields $C_{m\, ij}$ of the 3-form. These transform as
\begin{align}
\delta V &= 
	\mathscr L_\tau V 
	+ \tfrac1{2i}\left(\Lambda - \bar \Lambda \right)
	- \partial U ~.
\end{align}		
Their chiral field strengths $W_{\alpha ij}$ 
\begin{align}
W_\alpha &= -\tfrac14 \bar \nabla^2 \nabla_\alpha V 
	+\partial \Sigma_\alpha 
	+ \iota_{\mathcal W_\alpha} \Phi
\end{align}
should not be confused with those of the KK vectors. 
	\item[2-forms] There are seven 2-form gauge fields $C_{m n \, i}$ in the chiral spinor superfields $\Sigma_{\alpha i}$
\begin{align}
\delta \Sigma_\alpha &= 
	\mathscr L_\tau \Sigma_\alpha 
	-\tfrac14 \bar {\nabla}^2 {\nabla}_\alpha U 
	+ \partial \Upsilon_\alpha 
	+ \iota_{\mathcal W_\alpha} \Lambda~.
\end{align}
The field strength $H_i$ is real
\begin{align}
H &= \tfrac1{2i}\left(\nabla^\alpha \Sigma_\alpha - \bar \nabla_{\dalpha} \bar \Sigma^{\dalpha} \right)
	-\partial X 
	-\omega_{\mathsf h}(\mathcal W, V)
\end{align}
where we define the shorthand
\begin{align}
\omega_{\mathsf h}(\chi_\alpha, v) &:= 
\iota_{\chi^\alpha} {\nabla}_\alpha v 
	+ \iota_{\bar \chi_{\dalpha}} \bar {\nabla}^{\dalpha} v 
	+\tfrac12 \left(  \iota_{{\nabla}^\alpha\chi_\alpha} v 
		+ \iota_{\bar {\nabla}_{\dalpha} \bar \chi^{\dalpha}} v 
	\right)
.
\end{align}
It contains an excess of seven real scalars in the bottom component of $H_i$, but these
will be shown to be pure gauge.
	\item[3-form] The 3-form $C_{mnp}$ is encoded in a real scalar superfield $X$ with gauge transformation
\begin{align}
\delta X &= 
	\mathscr L_\tau X
	+\tfrac1{2i}\left({\nabla}^\alpha \Upsilon_\alpha
		- \bar {\nabla}_{\dalpha} \bar \Upsilon^{\dalpha} \right)
	- \omega_{\mathsf h}(\mathcal W_\alpha, U) 
\end{align}
and a ``reduced'' chiral field strength 
\begin{align}
G&= -\tfrac14 \bar \nabla^2 X 
	+ \iota_{\mathcal W^\alpha} \Sigma_\alpha
\end{align}
The bottom component of $G$ is a complex scalar, which can be interpreted as a the chiral compensator superfield of (modified) old minimal supergravity \cite{Grisaru:1981xm, Gates:1980az} (also known as 3-form supergravity \cite{Ovrut:1997ur})]. Its phase can be eliminated by a choice of $U(1)_R$ gauge and its modulus can be eliminated (or absorbed into the metric) by a choice of Weyl gauge.
\end{description}

Being given explicitly in terms of the prepotential superfields, the field strengths are the solutions to the Bianchi identities
\begin{subequations}
\label{E:BI}
\begin{align}
0 &= - \partial E
	\\
\tfrac1{2i}\left(E - \bar E\right) &=  \partial F
	\\
-\tfrac14 \bar {\nabla}^2 {\nabla}_\alpha F  &= 
	-\partial W_\alpha 
	- \iota_{\mathcal W_\alpha} E
	\\
\label{E:NATHBI1W}
\tfrac1{2i}\left({\nabla}^\alpha W_\alpha - \bar {\nabla}_{\dalpha} \bar W^{\dalpha} \right)&= 
	\partial H 
	+ \omega_{\mathsf h}(\mathcal W, F)
	\\ 
-\tfrac14 \bar {\nabla}^2 H &= 
	-\partial G 
	- \iota_{\mathcal W^\alpha} W_\alpha 
	\\
\bar {\nabla}_{\dalpha} G &= 0.
\end{align}
\end{subequations}
expressing the fact that these forms are closed in the extended de Rham complex \cite{Becker:2017njd}.

These rules and definitions are compatible with local superconformal symmetry on $\bm X$. 
In particular, we can consistently assign scaling dimensions ($\Delta$) and $U(1)_R$ weights ($w$) in addition to engineering dimension ($d$) to all the gauge parameters, prepotentials, and field strengths.
These are summarized in table \ref{T:Superscale}.
\begin{table}[t]
\begin{align*}
{\renewcommand{\arraystretch}{1.3} 
\begin{array}{|c|ccccccc|cccccc|}
\hline
	& H^a & \Psi_\alpha & G(X) & H(\Sigma_\alpha) & W_\alpha(V) &  F(\Phi) & \mathcal W_\alpha(\mathcal V)
	& L_\alpha & \Xi_{\alpha i} & \Omega^i & \Upsilon_\alpha& U_i & \Lambda_{ij} \\
\hline
\Delta
	&  -1 & -\tfrac32 & 	3(2) 	& 2 (\tfrac32) & \tfrac32(0) & 0(0) & \tfrac32(0)
	& -\tfrac32& -\tfrac32   & -3 & \tfrac32 & 0 &  0  \\
w 		& 0 	& -1  &   2(0) 	&  0 (1) 	& 	1(0)	& 0(0)& 	1(0)
	&-1 		&-1 	& 2 & 1&	0& 0 \\
d 			& -1	&- \tfrac12  &	0(-1)	&	0(-\tfrac12)& \tfrac12(-1) & 0(0)&\tfrac12(-1)
	&-\tfrac32&-\tfrac12& -1 & -\tfrac32& -2& -1 \\
\hline
\end{array}
}
\end{align*}
\caption{Weyl ($\Delta$) weight, $U(1)_R$ ($w$) weight, and engineering dimension of various
fields and parameters.}
\label{T:Superscale}
\end{table}

\subsection{Chern-Simons and K\"ahler actions}
\label{S:Review:Action}
The Chern-Simons action associated with this 3-form hierarchy can be obtained by first constructing $F\wedge F$ in superspace. This was the approach taken in reference \cite{Becker:2017njd} where it was shown that the composite superforms are given by
\begin{subequations}
\begin{align}
\label{E:NACS2}
\mathbb F &= \left( E + \bar E\right) F
\\
\mathbb W_\alpha&= E W_\alpha -\tfrac i4 \bar {\nabla}^2 (  F {\nabla}_\alpha F) 
\\
\mathbb H &= \left( E + \bar E\right) H + \omega(W, F) 
	- i {\nabla}^\alpha F\iota_{\mathcal W_\alpha}F 
	+ i \bar {\nabla}_{\dalpha} F \iota_{\bar{\mathcal W}^{\dalpha} }F 
\\
\mathbb G&= E G + \tfrac12 W^\alpha W_\alpha - \tfrac i4 \bar {\nabla}^2 (  F H) ~.
\end{align}
\end{subequations}
These complicated composite fields satisfy the same Bianchi identities (\ref{E:BI}) as the field strengths after which they are named. 
(This apparently highly non-trivial fact is simply the fact that the wedge product of closed forms is closed \cite{Becker:2017njd}.)
Because of these closure relations, the invariant action can be written
\begin{align}
\label{E:NACS}
S_{CS} &= {1\over\kappa^ 2}\int d^4 x \int_Y L_{CS}~, \eol
-12 L_{CS} &= 
	i \int d^2 \theta  \, \mathcal E\,   \left[  \Phi \mathbb G + \Sigma^\alpha \mathbb W_\alpha
	\right]
	+ \int d^4 \theta  \, E\,  \left[ V \mathbb H - X \mathbb F
	\right]
	+\mathrm{h.c.} 	
\end{align}
Here $E$ and $\cE$ are the full superspace and chiral superspace volume densities, respectively.\footnote{The volume densities are further discussed in Appendix \ref{A:Geometry}.
They are $y$-independent as they are built out of the supervielbein,
which is assumed $y$-independent as discussed in section \ref{S:Review:Metric}. This ensures gauge invariance under internal diffeomorphisms.}
The superconformal weights of these measures can be found in table \ref{T:ActionWeights}.
\begin{table}[t]
\begin{align*}
{\renewcommand{\arraystretch}{1.3} 
\begin{array}{|c|ccccccc|}
\hline
	&\kappa^2& \int d^7y & \int\! E & \int \!\cE & \nabla_{\alpha} & \nabla_a & \partial_i \\
\hline
\Delta &0& 0& -2 & -3 & \tfrac{1}{2} & 1 & 0 \\
w &0& 0&0  &-2 & -1 & 0 & 0 \\
d  &-9& -7 & -2& -3 & \tfrac{1}{2} & 1& 1\\
\hline
\end{array}
}
\end{align*}
\begin{caption}{
Weyl ($\Delta$) and $U(1)_R$ ($w$) weights of various measures and actions.
We denote $\int \!E = \int d^4x\, d^4\theta\, E$ and $\int \!\cE = \int d^4x\, d^2\theta\, \cE$.
}
\label{T:ActionWeights}
\end{caption}
\end{table}

In addition to the Chern-Simons action, one can construct gauge-invariant actions built purely from the curvatures themselves. 
The most general possibility is given by \cite{Becker:2017zwe}
\begin{align}
\label{E:Kahler}
S_{K} = -\frac3{\kappa^2} \int d^4 x \int d^7y \int d^4\theta \,E  \, \sqrt{g(F)}  \, 
	(\bar GG)^{1/3} \,{\mathcal F}(x) 
.	
\end{align}
The ingredients are as follows (cf.\ \S{}\ref{S:Review:3form}):
\begin{itemize}
	\item $g(F) = \mathrm{det}(g_{ij}(F))$ is the determinant of the Riemannian metric on $Y$ obtained from $F_{ijk}$ using \eqref{E:metric}. This factor is needed for the integrand to be a scalar density under internal diffeomorphisms \cite{Hitchin:2000jd, Hitchin2001,Becker:2016edk}.
	\item $G$ is the superconformal primary chiral superfield strength carrying the 4-form field strength $F_{mnpq}$ along the four-dimensional spacetime. 
It has weights $(3,2,1)$ (cf.\ table \ref{T:Superscale}) and is the compensator of (modified) old minimal (or 3-form) 4D $N=1$ supergravity. Its presence ensures the proper conformal weight of the integrand.
	\item $\cF$ is (for now) an arbitrary analytic function of the Weyl-invariant combination 
\begin{align}
\label{E:x}
x := |H|^2 = (\bar GG)^{-2/3} \,g^{ij} H_i H_j~
\end{align}
of the 3-form field strengths $F_{mnp \, i}$. Its precise definition will be given later.
\end{itemize}
We will refer to $S_K$ as the \emph{K\"ahler action}.\footnote{Strictly speaking, this is a misnomer since a K\"ahler potential depends on chiral multiplets and vector multiplets gauging their isometries. The deformation considered here also contains the tensor multiplets (chiral spinors) $\Sigma_{\alpha i}$.} One of our goals of this paper is to determine $\cF$ from invariance under the extended (non-manifest) supersymmetry (cf.\ \S{}\ref{S:main}).

\subsection{Additional gravitino superfields}
\label{S:Review:Grino}

In addition to the above supergravity-coupled action, we must introduce gravitino superfields that capture the dynamics of the seven additional spin-3/2 fields and which make the full action invariant under extended supersymmetry. These should be contained within an unconstrained spinor superfield $\Psi_{\alpha i}$ subject to a large gauge transformation that eliminates most of its component fields. The goal is to determine that gauge transformation, both for the gravitini and for the matter fields, and determine what constraints it places on the function $\cF(x)$.

In \cite{Becker:2017zwe}, we constructed the second-order action for fluctuations about a flat Minkowski background times a $G_2$-holonomy manifold $Y$ with closed and co-closed 3-form $\varphi_{ijk}(y)$ and its corresponding $G_2$-holonomy metric $g_{ij}(y)$. Two prepotentials took background values: $\vev\Phi = i \varphi$ and $\vev X = \theta^2$.
The fluctuation fields and the gravitino superfield were subject to the linearized gravitino gauge transformations
\begin{subequations}\label{E:Grinolin}
\begin{align}
\delta_0 \Psi_{\alpha i} &= \Xi_{\alpha i} + g_{i j} D_\alpha \Omega^j  \label{E:Grinolin.a}
\\
\delta_1 \Phi_{ijk} &= \tfrac1{2i} \tilde \varphi_{ijkl} \bar D^2 \bar \Omega^l
\\
\delta_1  V_{ij} &= \tfrac1{2i} \varphi_{ijk} (\Omega^k -\bar \Omega^k)
\\
\delta_1  \Sigma_{\alpha i}&= -\Xi_{\alpha i} \label{E:Grinolin.d}
\\
\delta_1 {\mathcal V}{}^i &= -\tfrac12(\Omega^i + \bar \Omega^i)~.
\end{align}
\end{subequations}
The field $\Xi_{\alpha i}$ is a chiral spinor and $\Omega^i$ is an unconstrained complex superfield. Here we have assigned a gravitino weight 1 to $\Psi$, $\Xi$, and $\Omega$, and denoted the above transformations by $\delta_0$ and $\delta_1$ corresponding to how they change the gravitino weight of the corresponding field.

One expects on general grounds that the full non-linear action takes the form of a power series expansion in the gravitino superfield $\Psi_{\alpha i}$. Schematically,
\begin{align}
\label{E:PertAction}
S_\textrm{11D} = S_0 + S_1 + S_2 + O(\Psi^3)~,
\end{align}
where $S_0$ consists of the zeroth-order K\"ahler \eqref{E:Kahler} and Chern-Simons actions \eqref{E:NACS}. The extended supersymmetry transformations should act on the superfields schematically as
\begin{align}
\delta = \delta_0 + \delta_1 + \delta_2 + \dots 
\end{align}
where $\delta_0$ is non-vanishing only for the gravitino and corresponds to the non-linear generalization of \eqref{E:Grinolin.a}. Then solving the equations $\delta S_\textrm{11D} = 0$ order-by-order, that is,
\begin{align}
\delta_1 S_0 + \delta_0 S_1 = 0~, \qquad
\delta_2 S_0 + \delta_1 S_1 + \delta_0 S_2 = 0~, \qquad \text{etc.}
\end{align}
we can determine the higher-order modifications of the action as well as higher-order modifications to the supersymmetry transformations. Note that these transformations are infinitesimal (thus first-order) in $\Xi$ and $\Omega$, and so $\delta_n \Psi$ will be $\cO(\Psi^{n})$ while for any other field $\delta_n$ will be $\cO(\Psi^{n-1})$.

In this paper, we will be concerned with the first step of the Noether procedure, that is, determining the complete first-order modification $S_1$ to the action and the full contributions to $\delta_1$ for the matter fields and $\delta_0$ for the gravitino superfield. The complete expressions for these quantities are given in eqs. \eqref{E:DeltaPot}, \eqref{E:DeltaGrino}, and \eqref{E:S1} of the next section.

\section{Extended supersymmetry and the gravitino supercurrent}
\label{S:GravitinoSupercurrent}

We begin this section by summarizing our main result: the non-linear extensions to the linearized transformation rules \eqref{E:Grinolin} and the correction to the action to first order in gravitino weight. Subsequently, we will motivate these results, discuss their underlying physics, and sketch some of the derivations.

\subsection{The main result}
\label{S:main}
As we are employing a covariant framework for the Kaluza-Klein gauge prepotential $\cV^i$, it appears implicitly in almost all of our formulae. For example, the spinor covariant derivatives
$\nabla_\alpha$ and $\bar\nabla_\dalpha$ are given as
\begin{align}
\nabla_\alpha = e^{i \cL_\cV} \nabla^0_\alpha e^{-i \cL_\cV} 
	= \nabla_\alpha^0 - \cL_{A_\alpha}~, \qquad
\bar\nabla_\dalpha = e^{-i \cL_\cV} \bar\nabla^0_\dalpha e^{i \cL_\cV} 
	= \bar\nabla_\dalpha^0 - \cL_{\bar A_\dalpha}~,
\end{align}
where $\nabla^0$ does not possess the KK connection. This means that covariantly
chiral superfields such as $\Phi$ must be understood as $\Phi = e^{-i \cL_\cV} \Phi^0$
where $\Phi^0$ is chiral with respect to $\nabla^0$. Then the variation of $\Phi$
has two parts arising from the variation of the independent constituents
$\Phi^0$ and $\cV$, so that
\begin{align}
\delta \Phi = \Delta \Phi - i \cL_{\delta \cV} \Phi
\end{align}
where $\Delta\Phi$ is chiral with respect to $\nabla$. We call $\Delta \Phi$
the \emph{covariantized transformation} of $\Phi$.
Similar comments pertain to $\Sigma_\alpha$; on top of this, it is convenient to introduce
additional pieces for its covariantized transformation. Similarly, additional terms are
naturally included in the covariantized transformations of the other superfields. We find
\begin{subequations}  \label{E:DeltaPotDef}
\begin{align}
\Delta \Phi &:= \delta \Phi + i \cL_{\delta \cV} \Phi~, \\
\Delta V &:= \delta V + \frac{1}{2} \iota_{\delta\cV} (\Phi + \bar \Phi)~,\\
\Delta \Sigma_\alpha
	&:= \delta \Sigma_{\alpha}
	+ i \cL_{\delta \cV} \Sigma_\alpha
	+ \frac{i}{4} \bar\nabla^2 \nabla_\alpha \big(\iota_{\delta\cV} V \big)
	- \frac{i}{2} \bar\nabla^2 \big(\iota_{\delta\cV} \nabla_\alpha V \big)~,\\
\Delta X &:= \delta X 
	+ \nabla^\alpha (\iota_{\delta\cV} \Sigma_\alpha) 
	+ \bar\nabla_\dalpha (\iota_{\delta\cV} \bar\Sigma^\dalpha) 
	- \frac{1}{2} \iota_{\delta\cV} (\nabla^\alpha \Sigma_\alpha + \bar\nabla_\dalpha \bar\Sigma^\dalpha)
	\cr & \quad
	- i \nabla^\alpha (\iota_{\delta\cV} \,\iota_{\cW_\alpha} V)
	+ i \bar\nabla_\dalpha (\iota_{\delta\cV} \,\iota_{\bar\cW^\dalpha} V)
	+ i\, \iota_{\delta\cV} \,\iota_{\cW^\alpha} \nabla_\alpha V
	- i\, \iota_{\delta\cV} \,\iota_{\bar\cW_\dalpha} \bar\nabla^\dalpha V
\end{align}
\end{subequations}
for the prepotentials of the tensor hierarchy. We emphasize that $\Delta \Phi$
and $\Delta \Sigma_\alpha$ are both chiral with respect to $\nabla$. 
(Additional comments can be found in Appendix \ref{A:CovTfor}.)

In terms of these covariantized transformations, the first-order extended SUSY transformations are
\begin{subequations}
\label{E:DeltaPot}
\begin{align}
\Delta_1 \Phi &= \Xi^\alpha \wedge W_\alpha
	+ \bar\nabla^2 \Big(
	\frac{i}{2} \iota_{\bar\Omega} \cU
	+ \frac{1}{4} \iota_{\bar\Omega} (F \wedge H)
	\Big)
	\label{E:DeltaPot.a}
	~,\\
\Delta_1 V &= - \frac{i}{2} \bar G \iota_\Omega F + \frac{i}{2} G \iota_{\bar\Omega} F ~,
\label{E:DeltaPot.b} \\
\Delta_1 \Sigma_\alpha &= -G\, \Xi_{\alpha}~, 
\label{E:DeltaPot.c} \\
\Delta_1 X &= -\frac{i}{2} \bar G\, \iota_\Omega H + \frac{i}{2} G\, \iota_{\bar\Omega} H~,
\label{E:DeltaPot.d} \\
\delta_1 \cV &= -\frac{1}{2} \bar G\, \Omega - \frac{1}{2} G\, \bar \Omega ~.
\label{E:DeltaPot.e}
\end{align}
\end{subequations}
In the transformation rule for $\Phi$, we have introduced the composite 4-form
\begin{align}\label{E:DefU}
\cU_{ijkl} &:= 3 \,\epsilon_{ijklmnp} \frac{\pa}{\pa F_{mnp}} \Big( \sqrt g\, \cF\, (G \bar G)^{1/3}\Big) \cr
	&= \frac{1}{3!} \sqrt{g}\,\epsilon_{ijklmnp} \Big(
	F^{mnp} (\cF + |H|^2 \cF')
	- 9 \cF'\, H^{m} F^{n p q} H_q (G \bar G)^{-2/3}
	\Big) (G \bar G)^{1/3}~.
\end{align}
The antisymmetric symbol $\epsilon_{ijklmnp}$ is a tensor density with entries $\pm1$, and upper indices in the second line of \eqref{E:DefU} have been raised with the metric $g_{ij}(F)$.

As we will explain presently, the zeroth-order transformation of the gravitino superfield turns out to be
\begin{align}
\label{E:DeltaGrino}
\delta_0 \Psi_{\alpha i}
	&= \Xi_{\alpha i}
	+ \bar G\, \cG_{i j} \nabla_\alpha \Omega^j
	+ i \,W_{\alpha i j} \,\bar{\Omega}^{j}
~,
\end{align}
where $\Xi_{\alpha i}$ is covariantly chiral and $\Omega^i$ is an unconstrained complex superfield, as in the linearized theory. In the full theory, we see that the second term is dressed with a factor of $G$ as well as a complex rank-2 tensor $\cG_{i j}$ (see eq. \eqref{E:DefGij} for its definition) and a third term involving the 2-form field strength $W_{\alpha i j}$ has appeared. The first-order gravitino contribution to the action is
\begin{align}
\label{E:S1}
S_{1} = \frac{1}{\kappa^2} \int d^4 x \int d^7y \int d^4\theta \,E  \,
	(\Psi_i^\alpha J_\alpha{}^i + \text{c.c.})
~,
\end{align}
where the complex superfield $J_\alpha{}^i$ is given by
\begin{align}
\label{E:DefJ.v1}
J_\alpha{}^i
	&=
 	-  \sqrt{g} \, (G \bar G)^{1/3}\, (\cF - 2 |H|^2 \cF')\, \cW_\alpha{}^i
	- 3 i \, G \,\nabla_\alpha \Big(
		(G \bar G)^{-1/3} \sqrt g\, \cF'\, g^{i j} H_j \Big)
	\cr & \quad
	- \frac{i}{96} \epsilon^{ijklmnp} \,\cU_{jklm}\,W_{\alpha n p}
	+ \frac{1}{144} \epsilon^{ijklmnp} F_{jkl} \Big(G \,\nabla_\alpha F_{mnp}
	- 3 \,H_m W_{\alpha np} \Big)
~.
\end{align}
Since $\Psi$ gauges the seven supersymmetries not manifest in $N=1$ superspace, $J_\alpha{}^i$ can be interpreted as the Noether supercurrent for these supersymmetries. 
One can easily check that the first line counters the $\Xi$ transformation of the K\"ahler term whereas the second line counters the $\Xi$ transformation of the Chern-Simons term.

Further requiring $\Omega$-invariance generates no new contributions to the supercurrent, but instead determines the function $\cF(x)$ (recall eq.\ \ref{E:x}) in terms of
\begin{align}
\hat \cF := \cF - 2 x \cF'
\end{align}
as the solution to the quartic polynomial
\begin{align}\label{E:DefF}
\frac{x}{4} \hat \cF^4 + \hat \cF^3 - 1 = 0~.
\end{align}
Together these imply, for example, that
$\cF'(x) = \frac{1}{12} \hat \cF^2$ and
$\cF(x) = \frac{1}{3} \hat\cF + \frac{2}{3} \hat \cF^{-2}$, so that
both $\cF$ and its first derivative can be expressed in terms of $\hat \cF$.
Both $\cF(x)$ and $\hat\cF(x)$ possess infinite series expansions:
\begin{align}
\cF(x) = 1 + \frac{x}{12} - \frac{x^2}{144} + \cdots~, \qquad
\hat \cF(x) = 1 - \frac{x}{12} + \frac{x^2}{48} + \cdots
\end{align}
Both functions are monotonic. Whereas $\cF(x)$ slowly increases without bound, $\hat \cF(x)$ slowly tends to zero.

Finally, the function $\cG_{i j}$ in the gravitino transformation \eqref{E:DeltaGrino} is given by
\begin{align}\label{E:DefGij}
\cG_{i j} 
	= (G \bar G)^{-1/3} \Big(\hat \cF^{-1}g_{i j} + \frac i2 \hat \cF F_{i j k} \,g^{k l} \,H_l (G \bar G)^{-1/3} \Big)
~.
\end{align}
It has $(\Delta, w , d) = (-2, 0, 0)$ and is a Hermitian matrix: The first term is real and symmetric, and the second term is imaginary and antisymmetric.
These results appear rather complicated, but there proves to be a great deal of structure that constrains them. 
In the remainder of this section, we will motivate these results and sketch their derivations.

\subsection{The $\Xi$-transformations and the gauge-for-gauge symmetry of $\Omega$}
Let us begin by justifying the $\Xi$ transformations. We have chosen to fix the gravitino's conformal and $U(1)_R$ weights as in table \ref{T:Superscale}, and to identify $\Xi_{\alpha i}$ with no additional factors as in \eqref{E:DeltaGrino}. Taking into account the engineering and conformal dimensions of the various fields, it is easy to see that we cannot assign a $\Xi$ transformation to $V_{ij}$, $X$, or $\cV^i$, provided we expect only field strengths to appear on the right-hand-side of the covariantized transformations.

Now keeping in mind the linearized transformation rule \eqref{E:Grinolin.d} for $\Sigma_{\alpha i}$, the only non-linear modification consistent with chirality involves a factor of $G$ correcting the conformal and $U(1)_R$ weights. Incidentally, this is also why it is necessary to choose $\Xi_{\alpha i}$ to have a lower $GL(7)$ index. If the index were raised, we would need to lower it with a chiral metric, but no chiral metric exists.

Considerations of various weights and chirality similarly restrict the $\Xi$ transformation of the remaining prepotential $\Phi_{ijk}$ to at most be of the form \eqref{E:DeltaPot.a}. The coefficient can be determined by requiring $\Xi$-invariance of the terms in the Chern-Simons action that are purely chiral -- that is, terms that cannot be lifted to full superspace by eliminating an overall $\bar \nabla^2$ factor. Because the supercurrent term \eqref{E:S1} is a full superspace integral, the purely chiral terms must cancel on their own, and this is only possible if $\Phi$ transforms as in \eqref{E:DeltaPot.a}. (To verify this, it helps to use the expression \eqref{E:deltaSCS} for the general variation of the Chern-Simons action.)

The $\Xi$ transformations now uniquely determine the first-order gravitino action \eqref{E:S1}. Because these transformations can be written purely in the language of differential forms, knowing nothing of the metric $g_{ij}$ or the function $\cF$, the K\"ahler and Chern-Simons actions must be canceled separately by their respective gravitino terms. The first line of \eqref{E:DefJ.v1} exactly cancels the transformations of $\Phi_{ijk}$ and $\Sigma_{\alpha i}$ in the K\"ahler term, whereas the second line of \eqref{E:DefJ.v1} can be seen to cancel the Chern-Simons term. The only terms in $S_1$ we might not determine in this way are those that are $\Xi$-invariant -- that is, terms where the gravitino appears only as $\bar\nabla_\dalpha \Psi_{\alpha i}$. But no such terms can be written down by virtue of conformal and engineering dimension arguments.

Similar arguments can be made to motivate the $\Omega$ transformations. There is no transformation one can postulate for $\Sigma_{\alpha i}$ that is consistent both with chirality and its various weights. The other prepotentials are more subtle. Let us focus first on $\cV^i$. The transformation given in \eqref{E:DeltaPot.e} matches in the linearized approximation and is the only non-linear possibility up to multiplication by an overall function of the weightless combination $|H|^2$ \eqref{E:x}. To eliminate this possibility requires some physical insight. At the linearized level, the gravitino superfield transformation \eqref{E:Grinolin.a} is subject to a gauge-for-gauge symmetry whereby $\Omega$ can be shifted by an antichiral superfield. (This is necessary so that the counting of the degrees of freedom is correct: an $\Omega$ without such a shift symmetry would introduce an additional gauge invariance absent in the linearized theory.) The transformation \eqref{E:DeltaPot.e} respects this gauge-for-gauge symmetry, but any such $|H|^2$ modifications would break it. That is, if $\Omega^i$ is an antichiral superfield $\bar\phi^i$, the transformation \eqref{E:DeltaPot.e} leads to
$\delta_1 \cV^i = -\frac{1}{2} (\bar G \bar\phi^i + \text{h.c.})$,
which is the form of a chiral gauge transformation of $\cV$,
$\delta \cV^i = \lambda^i + \bar \lambda^i$, see section 3.1 of \cite{Becker:2017zwe}.
A compensating chiral gauge
transformation with $\lambda^i = \frac{1}{2} G \phi^i$ leaves $\cV$ invariant.
A similar argument fixes the transformations of $V_{i j}$ and $X$.

To fix the $\Omega$-transformation for $\Phi_{ijk}$ and to uncover the full expression for the $\Omega$-transformation of the gravitino is more subtle. A practical approach is to first require $\Omega$ invariance around a field configuration with $H_i = 0$. An obvious ansatz here is to generalize the linearized results to (introducing factors of $G$ for weight)
\begin{subequations}\label{E:DeltaH=0}
\begin{align}
\delta_\Omega \Psi_{\alpha i}
	&= g_{i j} \,\bar G (G \bar G)^{-1/3} \nabla_\alpha \Omega^j + \cdots~,
	\label{E:DeltaH=0.a} \\
\Delta_\Omega \Phi_{ijk} &= -\frac{i}{2} 
	\bar\nabla^2 \Big((G \bar G)^{1/3} \tilde F_{ijkl} \,\bar\Omega^l\Big)
	+ \cdots \label{E:DeltaH=0.b}
\end{align}
\end{subequations}
This $\Phi$ transformation in the $\Phi W^\alpha W_\alpha$ part of the Chern-Simons action \eqref{E:NACS} 
can only be canceled if $\delta\Psi_{\alpha i}$ acquires the term $i \,W_{\alpha ij} \bar\Omega^j$. 
(Here the results of Appendix \ref{A:VaryExp} are useful.) 
Thus, 
\begin{align}
\delta_\Omega \Psi_{\alpha i}
	&= g_{i j} \,\bar G (G \bar G)^{-1/3} \nabla_\alpha \Omega^j + i \,W_{\alpha ij} \bar\Omega^j + \cO(H)~.
\end{align}
The $\bar\Omega$ term does not actually violate the gauge-for-gauge symmetry, as one can counter the chiral shift in $\bar \Omega$ by a shift of $\Xi_{\alpha i}$ involving $W_{\alpha i j}$. In order for this chiral shift symmetry to hold at higher order in $H_i$, the 
$i \,W_{\alpha ij} \bar\Omega^j$ term that we have added cannot be dressed by any function of $|H|^2$. So the most general expectation for the gravitino is the expression \eqref{E:DeltaGrino} for some complex rank-2 tensor $\cG_{i j}$ agreeing with $(G \bar G)^{-1/3} g_{ij}$ when $H_i$ vanishes. (The exact expression \eqref{E:DefGij} cannot be so simply determined.)
Let us record here the explicit form for the gauge-for-gauge symmetry:
\begin{align}
\Omega^i = \bar\phi^i~, \qquad
\lambda^i = \frac{1}{2} G \phi^i~, \qquad 
\Xi_{\alpha i} =  -i W_{\alpha i j} \phi^j ~, \qquad \bar\nabla_\dalpha \phi^i = 0~.
\end{align}

We now have sufficient information to completely determine the $\Phi_{ijk}$ transformation to all orders in $H_i$. Let's assume that the full transformation of $\Phi$ is
\begin{align}
\Delta \Phi_{ijk} &= -\frac{i}{2} 
	\bar\nabla^2 \Big(Z_{ijkl} \,\bar\Omega^l
	\Big)
	+ 3 \,\Xi^\alpha_{[i} W_{\alpha jk]}
\end{align}
for some unknown covariant tensor $Z_{ijkl}$. This must reduce to \eqref{E:DeltaH=0.b} in the limit where $H$ vanishes.
Now under the gauge-for-gauge symmetry (including the appropriate non-abelian transformation), we find
\begin{align}\label{E:DeltaPhiTemp}
\Delta \Phi_{ijk} &=
	- \frac{i}{2} \phi^l \bar\nabla^2 Z_{ijkl}
	+ 3 i \,\phi^{l}  W^\alpha{}_{[l i} W_{\alpha jk]}
	+ 4 i \, \phi^{l} \pa_{[l} \Phi_{ijk]}~.
\end{align}
This cannot be made to vanish since one cannot choose a covariant $Z_{ijkl}$ to cancel the other two terms. What one can do instead is to arrange for $\Delta \Phi_{ijk}$ to generate a \emph{trivial symmetry} of the action by taking
\begin{align}
\Delta \Phi_{i j k} \propto \epsilon_{ijklmnp} \,\phi^{l} \frac{\delta S_0}{\delta \Phi_{mnp}}~,
\end{align}
so that $\delta S_0 = 0$ \emph{automatically}. This requires
\begin{align}
Z_{ijkl}
	&= 3\, \epsilon_{ijklmnp} \frac{\pa}{\pa F_{mnp}}(\sqrt g\, (G \bar G)^{1/3} \cF)
	-2 i \, F_{[ijk} H_{l]}~.
\end{align}
The first term is the variation of the K\"ahler term with respect to $\Phi$.
The $F \wedge H$ term combines with the $W^\alpha \wedge W_\alpha$ and $\pa \Phi$ terms in \eqref{E:DeltaPhiTemp} to give the variation of Chern-Simons action with respect to $\Phi$. This leads to the full expression \eqref{E:DeltaPot.a}, and agrees (as it must) with the result when $H_i$ vanishes.

\subsection{Determining $\cF$ and $\cG_{i j}$ from $\Omega$ transformations}

We have not yet justified the forms of $\cF$ and $\cG_{i j}$. These will come from requiring invariance under $\Omega$ transformations, but we can already make a few comments about $\cG_{i j}$. Because it can only be built from $F_{ijk}$, $G$, and $H_i$ (no derivatives may appear if the action should be two-derivative),
\begin{align}
\cG_{i j} = (G \bar G)^{-1/3} \Big(c_0(x) g_{i j} + c_1(x) F_{i j k} g^{k l} H_l (G \bar G)^{-1/3}
	+ c_2(x) H_i H_j (G \bar G)^{-2/3}\Big)
\end{align}
where the form factors $c_i$ are complex functions of $|H|^2$ \eqref{E:x}. (The factors of $G \bar G$ give the correct Weyl weights.) Spacetime parity further constrains $c_0$ and $c_2$ to be real and $c_1$ to be imaginary.

Deriving the form factors and the expression for $\cF$ is a relatively straightforward if laborious task. A number of ingredients aid in this. Collected in Appendix \ref{App:VaryCSK} are the general variations of the zeroth-order actions. The variation of the Chern-Simons action is naturally written as \eqref{E:deltaSCS} in terms of the covariantized variations of the potentials \eqref{E:DeltaPot}. The K\"ahler term is more naturally written as \eqref{E:deltaSK} in terms of variations of the field strengths themselves. These are derived from the variations of the prepotentials (including the Kaluza-Klein prepotential) and (focusing only on the $\Omega$ terms) are given by
\begin{subequations}\label{E:deltaFS}
\begin{align}
\delta_1 F
	&= 
	\frac{i}{2} \cL_{\bar G \Omega} F
	+ \frac{1}{4} \nabla^2 \Big((G\bar G)^{1/3} \iota_\Omega \cU \Big)
	+ \frac{i}{8} \iota_{\nabla^2 \Omega} (F \wedge H)
	+ \frac{i}{4} \iota_{\nabla^\alpha \Omega} \nabla_\alpha (F \wedge H)
	\cr & \quad
	+ \frac{1}{2} \iota_\Omega \,\Big(
		\pa \bar \Phi \bar G + \frac{i}{4} \nabla^2 (F \wedge H)
		\Big)
	+ \HC~, \label{E:deltaFS.F} \\
\delta_1 G
	&=
	\frac{i}{2} (\cL_{\bar G \Omega} - \cL_{G\bar \Omega}) G
	- \frac{i}{4} \bar\nabla^2 \bar\Omega^i H_i\, G
	- \frac{i}{2} \bar\nabla_\dalpha \bar\Omega^i \bar\nabla^\dalpha H_i\, G
	- i \,\iota_{\bar\Omega} \iota_{\cW^\alpha} W_\alpha\, G~, \label{E:deltaFS.G} \\
\delta_1 H
	&= \frac{i}{2} \cL_{\bar G \Omega} H
	+ \frac{1}{2} \bar G \,\iota_{\nabla^\alpha\Omega} (W_\alpha - 2 i \, \iota_{\cW_\alpha} F)
	+ i \bar G \,\iota_{\Omega} \iota_{\cW_\dalpha} \bar\nabla^\dalpha F
	\cr & \quad
	+ \frac{1}{2} \bar\nabla_\dalpha (\bar G \iota_\Omega \bar W^\dalpha)
	+ \HC \label{E:deltaFS.H}
\end{align}
\end{subequations}
An analogous expression can be given for $W_{\alpha i j}$, but it does not appear in the K\"ahler action so it is not as useful. In casting the result in this form, we have pushed derivatives onto $\Omega$ as much as possible. Note that the first term in each expression \eqref{E:deltaFS} corresponds to a uniform internal diffeomorphism, so this variation just leads in the K\"ahler action to a total internal derivative.

A tractable approach to checking $\Omega$ invariance (which we followed to completion) is to work order-by-order in $W_{\alpha ij}$ and $\cW_\alpha{}^i$. The terms quadratic in these field strengths cancel immediately, while the linear ones lead to the form factors \eqref{E:DefGij} 
\begin{align}
c_0 =\hat{\mathcal F}^{-1}
~,~~
c_1 = 6i \hat{\mathcal F}^{-1} {\mathcal F}' = \frac i2 \hat{\mathcal F}
~,~~
c_2 = 0
~,~~
\end{align}
and conditions \eqref{E:DefF}. The explicit calculation is not particularly enlightening, so we do not reproduce it here. Instead, in section \ref{S:CovariantForm} we will briefly sketch a more covariant approach that should lead to a simplification of this check and also permit computation of the gravitino terms to higher order.

Because of the complexity of the actions and transformation rules, it is easy to lose sight of an important fact: Under very mild assumptions about the basic structure of the $\Xi$ and $\Omega$ transformations, the complete first-order transformations of the matter fields and the gravitino have been determined, as well as the function $\cF$ in the K\"ahler term. It is clear that this should not be otherwise -- we are striving to describe 11D supergravity and that theory is unique -- but it is heartening to observe that the consistent Noether coupling of the additional seven gravitini does indeed determine the action and transformations uniquely. In the next section, we will show how to make the connection more transparent.

\section{Connecting with 11D superspace}
\label{S:11DConnection}

The $\Xi$ and $\Omega$ transformations associated with the extended supersymmetry of the gravitino superfields describe a rather large gauge group consisting of a chiral spinor $\Xi$ and an unconstrained complex superfield $\Omega$. 
As we have reviewed in \cite{Becker:2017zwe}, these transformations permit a Wess-Zumino gauge condition on $\Psi$ where only five components survive: the extended gravitini $\psi_{m \alpha i}$, a complex anti-self-dual antisymmetric tensor $t_{ab\, i}^-$, a complex vector $y_{m \,i}$, and a spinor $\rho_{\alpha i}$.\footnote{Actually, $\Psi$ contains only the gamma-traceless part of the extended gravitini. The spin-1/2 components are encoded in some of the fermions in the matter fields.} At the end of the day, two remnants of the superfield transformations survive: the extended supersymmetry transformation
$\xi_{\alpha i}$ corresponding to the bottom component of $\Xi_{\alpha i}\vert$ and a bosonic transformation with parameter $\nabla^2 \Omega^i\vert$, under which $y_{m \, i}$ transforms as a gauge field. Here let us denote this parameter $z^i$ and normalize it as the weight-less quantity
\begin{align}
z^i := \frac{i}{4} (G \bar G)^{1/3} \nabla^2 \Omega^i\vert~.
\end{align}

This bosonic gauge transformation played a critical role in arriving at the correct linearized action.
There are 16 extraneous scalar fields encoded in the $N=1$ superfields not present in 11D supergravity. 
Two of these (the bottom components of $G$) are Weyl and $U(1)_R$ compensators, which are natural from the point of view of $N=1$ superspace. The other 14 turned out to be  pure gauge degrees of freedom that could be removed by the bosonic gauge transformation involving $z^i$. While this leads to a consistent description of the underlying physics, it is slightly puzzling. After all, we could have imagined descending from eleven dimensions directly and defining our four dimensional fields. How would these 14 degrees of freedom appear in that dictionary? In this section, we will answer this question by descending directly from 11D superspace to 4D $N=1$ conformal superspace.

Eleven-dimensional superspace was introduced in \cite{Cremmer:1980ru, Brink:1980az} to describe (on-shell) 11D supergravity. (Our conventions here differ slightly to admit a closer connection to normalizations and conventions used in $4D$ $N=1$ superspace.) We denote 11D vector and spinor indices with hats. An 11D vector decomposes as $V^\ha = (V^a, V^{\ul a})$ and a 32-component Majorana spinor decomposes as $\Psi_\halpha = (\Psi_{\alpha I}, \Psi^{\dalpha I})$ where $\alpha$ and $\dalpha$ are two-component chiral and antichiral spinor indices for $SO(3,1)$, and $I$ denotes an $SO(7)$ spinor index, which can be raised or lowered with $\delta_{IJ}$. The 11D charge conjugation matrix is real and antisymmetric,
\begin{align}
C^{\halpha \hbeta} =
\begin{pmatrix}
-\eps^{\alpha\beta} \delta^{IJ} & 0 \\
0 & -\eps_{\dalpha \dbeta} \delta_{IJ}
\end{pmatrix}~.
\end{align}
Spinor indices are raised and lowered as $\Psi^\halpha = \Psi_\hbeta C^{\hbeta \alpha}$ and $\Psi_\halpha = \Psi^\hbeta C_{\hbeta \halpha}$. We take the $Spin(10,1)$ gamma matrices $(\Gamma_\ha){}_\halpha{}^\hbeta$ to be pseudo-Hermitian, obeying $\Gamma_\ha^\dag = \Gamma^\ha$. These decompose as
\begin{align}
(\Gamma^a)_\halpha{}^\hbeta =
\begin{pmatrix}
0 & i (\sigma^a)_{\alpha \dbeta} \delta_{IJ} \\
i (\bar\sigma^a)^{\dalpha \beta} \delta^{IJ} & 0
\end{pmatrix}~, \qquad
(\Gamma^{\ul a})_{\alpha}{}^\hbeta = 
\begin{pmatrix}
\delta_\alpha{}^\beta (\Gamma^{\ul a})_I{}^J & 0 \\
0 & -\delta^\dalpha{}_\dbeta (\Gamma^{\ul a})^I{}_J
\end{pmatrix}~,
\end{align}
where the $SO(7)$ gamma matrices $(\Gamma^{\ul a})_{I J}$ are imaginary and antisymmetric.

The superspace is described by a supervielbein $E_\hM{}^\hA$ and structure group connection $\Omega_\hM{}_\hA{}^\hB$ valued in $SO(10,1)$ so that $\Omega_\hM{}_\halpha{}^\hbeta = \frac{1}{4} \Omega_\hM{}^{\ha \hb} (\Gamma_{\ha \hb})_{\halpha}{}^\hbeta$ and $\Omega_\hM{}_\halpha{}^\hb = \Omega_\hM{}_\ha{}^\hbeta = 0$. The constraints on the torsion and curvature tensors imply that the geometry is completely on-shell with no auxiliary fields present. For example, the non-vanishing tangent-space components of the torsion tensor $T^\hA = \cD E^\hA$ are given by
\begin{align}
T_{\halpha \hbeta}{}^\hc &= 2 (\Gamma^\hc)_{\halpha \hbeta}~, \qquad
T_{\ha \hbeta}{}^\hgamma = -\frac{1}{36} F_{\ha\hb\hc\hd} (\Gamma^{\hb\hc\hd})_{\hbeta}{}^\hgamma
	- \frac{1}{288} (\Gamma_{\ha\hb\hc\hd\he})_{\beta}{}^\gamma F^{\hb\hc\hd\he}~,\cr
T_{\ha \hb}{}^\halpha &= -\frac{1}{84} (\Gamma^{\hc\hd})^\halpha{}^\hbeta \cD_\hbeta 
F_{\ha\hb\hc\hd}~,
\label{E:Torsion11D}
\end{align}
in terms of a superfield $F_{\ha\hb\hc\hd}$ which turns out to be the supercovariant 4-form field strength. As a consequence of the Bianchi identities, $F$ is covariantly closed.
The components of the super-Riemann tensor $R_{\hA \hB\, \hC}{}^\hD$ are also completely determined by the Bianchi identities, but we won't need them here. One can make the 3-form apparent in superspace as well by introducing a super 3-form $C_{\hM \hN \hP}$ with super 4-form field strength superfield $F = d C$. Aside from its top component, which must be the same superfield $F_{\ha \hb \hc \hd}$ appearing above, the only other non-vanishing component of $F$ is $F_{\halpha\hbeta \hc \hd} = 2 \,(\Gamma_{\hc \hd})_{\halpha\hbeta}$.
The full super 4-form is then given by
\begin{align}\label{E:11D.4form}
F &= - \frac{1}{2} E^\hc \wedge E^\hd \wedge E^\halpha \wedge E^\hbeta\, (\Gamma_{\hc \hd})_{\halpha \hbeta}
	+ \frac{1}{4!} E^\ha \wedge E^\hb \wedge E^\hc \wedge E^\hd \, F_{\hd \hc \hb \ha}~.
\end{align}
Closure is straightforward to check using \eqref{E:Torsion11D}.

Let us now descend to 4D $N=1$. The key question is how to identify the $N=1$ gravitino 1-form $E^\alpha$ in terms of the 11D 1-form $E^{\halpha} = (E^{\alpha I}, E_\dalpha{}^I)$. The most general possibility is
$E^{\alpha I} = \eta^I \,E^\alpha + \cdots$
where $\eta^I$ is some \emph{complex} $SO(7)$ spinor, which may depend on all coordinates, and the ellipsis denotes the other seven additional gravitino connections, which we will ignore. Actually, it is convenient to factor out a modulus $\phi$ from $\eta^I$ so that it is normalized to $\eta^I \bar \eta^I = 1$. 
Thus, we identify
\begin{align}
\label{E:11To4.Grav}
E^{\alpha I} \rightarrow \eta^I \,\phi^{1/2}\, E^{\alpha} + \cdots
~~~\textrm{and}~~~ 
E_\dalpha{}^{I} \rightarrow \bar\eta^I \,\phi^{1/2}\, E_{\dalpha} + \cdots~.
\end{align}
In order for the $N=1$ torsion tensor to be canonically normalized, we must introduce a factor of $\phi$ into $E^a$, that is 
\begin{align}\label{E:11To4.Viel}
E^\ha \rightarrow
\begin{pmatrix}
\phi E^a \\
E^{\ul a}
\end{pmatrix}~.
\end{align}
The superfield $\phi$ is a conformal compensator because it introduces a new Weyl symmetry under which $\phi$ has weight $1$, $E^a$ has weight $-1$, and $E^\alpha$ has weight $-1/2$. 
Given the constraint on the norm of $\eta^I$, any variation can be written as
\begin{align}\label{E:delta.eta}
\delta \eta^I = -i \omega \eta^I + y^i \, (\mathring\Gamma_{i})^{I J} \bar\eta^{J}
~~~\textrm{and}~~~
\delta \bar\eta^I = i \omega \bar\eta^I - \bar y^i \, (\mathring\Gamma_{i})^{I J} \eta^{J}~,
\end{align}
in terms of a complex $GL(7)$ vector $y^i$ and a real parameter $\omega$.
We have written the $SO(7)$ gamma matrices as $\mathring \Gamma_i = E_i{}^{\ul a} \Gamma_{\ul a}$, so that they possess a $GL(7)$ index. The 14 components of $y^i$ should evidently correspond to the complex gauge symmetry associated with $z^i$. The real parameter $\omega$ describes a local $U(1)_R$ transformation, because it can be absorbed by a phase rotation
$\delta E^\alpha = i \omega\, E^\alpha$.

We can flesh out these statements by constructing explicit expressions for the superfields $G$ and $H_i$. These are naturally encoded as the lowest components of their corresponding 4-form and 3-form field strengths in $N=1$ superspace \cite{Becker:2017njd}
\begin{align}
F_4 &= E^c \wedge E^d \wedge E^\alpha \wedge E^\beta \, (\sigma_{c d})_{\alpha\beta} \,\bar G 
	+ \HC + \cdots~, \cr
F_{3 i} &= E^c \wedge E^\alpha \wedge E^{\dbeta} \, i(\sigma_c)_{\alpha\dbeta} \,H_i + \cdots
\end{align}
These should arise from the decomposition of the 11D 4-form field strength \eqref{E:11D.4form} under the identifications \eqref{E:Torsion11D}, \eqref{E:11To4.Grav}, and \eqref{E:11To4.Viel}. The lowest dimension term of \eqref{E:11D.4form} exactly reproduces those above provided we identify
\begin{align}\label{E:11DefGH}
G = \bar \eta^2\, \phi^3\, ~, \qquad
\bar G = \eta^2 \, \phi^3\, ~, \qquad
H_i &= 2 \,(\eta \mathring\Gamma_i \bar \eta) \, \phi^2~.
\end{align}
Because we are being somewhat schematic with the reduction, we should probably not trust these results beyond lowest component. In particular, they may develop $\Psi$-dependent modifications. But this suggests that the bottom component of $H_i$ is associated with the complexity of $\eta$: If $\eta_I = \bar\eta_I$ (up to a phase), $H_i\vert$ would vanish. This is consistent with the interpretation of \cite{Becker:2016edk} that this field can be gauged away in the component theory.

We still need to identify how the internal metric coming from 11D supergravity, which we denote $\mathring{g}_{i j}$, is related to $g_{i j}(F)$. It is clear these \emph{cannot} be exactly the same because $g_{i j}(F)$ transforms under $z^i$ while $\mathring{g}_{i j}$ must be independent of $\eta$. Indeed, under the $z^i$ part of the $\Omega$ transformations we find
\begin{align}
\delta g_{i j} &= 
	- \frac{1}{3} g_{i j} \hH_{k} z^k
	+ \hH_{(i} g_{j) k} z^k
	- 6 i\,\cF'\, \hH_{(i} F_{j) k l} g^{k m} \hH_m z^l
	+ \HC~,\\
\delta \sqrt g &= -\frac{2}{3} \sqrt g\, \hH_i z^i + \HC
\end{align}
Here and below we use the Weyl-invariant combination $\hH_i := (G \bar G)^{-1/3} H_i$.
Whatever the internal metric $\mathring g_{i j}$ is, it must be invariant under $z^i$ transformations. To find it, the additional relations are useful (recall eq.\ \eqref{E:x}):
\begin{align}
\delta x = \delta |H|^2 = -4 \,(\hat \cF^{-1} + \tfrac{1}{3} x) \hH_{i} z^i + \HC~, \qquad
\delta \hat \cF = \frac{1}{3}\, \hat \cF\, \hH_i z^i + \HC
\end{align}
Using these results together with \eqref{E:11DefGH}, one can check that there is (up to normalization) only one symmetric rank-2 tensor that is $z$-invariant and Weyl-invariant. This should be identified with the internal 7D metric,
\begin{align}\label{E:11DefMetric}
\mathring g_{i j} &= \hat\cF g_{i j} + \frac{1}{4} \hat \cF^2\, \hH_i \hH_j~, \qquad
\det \mathring g = \det g\, \hat \cF^4~.
\end{align}
In addition to $\det \mathring g$, there is another scalar $z$-invariant,
$G^{1/3} \bar G^{1/3} \hat \cF^{-1}$, which carries Weyl-weight 2. The only such invariant scalar in the 11D theory is the conformal compensator, so we identify
\begin{align}
\phi^2 &= G^{1/3} \bar G^{1/3} \hat \cF^{-1} \quad \implies \quad \hat \cF = (\eta^2 \bar \eta^2)^{1/3}~.
\end{align}
This identifies $\hat \cF$ in terms of $\eta^2$. 
Now from the explicit equations \eqref{E:11DefGH} and \eqref{E:11DefMetric}, one can find that
\begin{align}
x := |\hH|^2 = 4 \hat \cF^{-4} - 4 \hat \cF^{-1}~.
\end{align}
This is nothing but the quartic polynomial \eqref{E:DefF}, here derived as an \emph{algebraic} equation when $x$ and $\hat \cF$ are both expressed in terms of $\eta^2$. 

To confirm these identifications, we should verify that one can consistently write down a relation between the parameters $z^i$ and $y^i$ in \eqref{E:delta.eta}. $G$ and $H_i$ vary under \eqref{E:delta.eta} as
\begin{align}
\label{E:deltayGH}
\delta G = 2 \phi^3 \,  \bar \eta \,\delta \bar\eta
	= \phi \, \bar y^i H_i, \quad
\delta H_i
	= 2 \phi^2\, \eta \mathring\Gamma_i \delta \bar \eta + \HC
	= -2 \phi^2\, y^j \, \eta^2 \mathring g_{i j} + \HC
\end{align}
We want to compare this to the transformation under $\delta_1$.
For $G$, this is straightforward from \eqref{E:deltaFS}, but for $H_i$ we must correct for the gravitino superfield. 
In the Wess-Zumino gauge (cf.\ app.\ C of \cite{Becker:2017zwe}) $\nabla^\alpha \Psi_{\alpha i}| \to 0$,  $\nabla^\alpha \Xi_{\alpha i}| = -  \bar G\, \cG_{i j}\nabla^2 \Omega^j| + \dots$ by \eqref{E:DeltaGrino}.
With this taken into account, we find
\begin{align}
\label{E:deltazGH}
\delta G = G\, (G \bar G)^{-1/3}\, \bar z^i H_i
~~~\textrm{and}~~~
\delta H_i = -2 (G \bar G)^{2/3} \cG_{i j} z^j + \HC
\end{align}
This can be identified with \eqref{E:deltayGH} if
\begin{align}\label{E:y.From.z}
y^i 
	&= \eta^2 \Big(
	\hat \cF^{-4}\, z^i
	- \frac{1}{4} \, g^{i j} \hH_j \hH_k z^k
	+ \frac{i}{2}\, \hat \cF^{-2} g^{i i'} g^{k k'} F_{i' j k'} \hH_k z^j
	\Big)~.
\end{align}

For additional confirmation, let us try to identify the bottom component of $F_{ijk}$. As with $H_i\vert$, this contains scalar fields not directly present in 11D: 28 of its 35 degrees of freedom arise from the internal metric via the $G_2$ relation \eqref{E:metric}, but 7 degrees of freedom remained. It turns out that $F_{ijk}$ has a remarkably simple interpretation expressed by
\begin{align}\label{E:11DefF}
F_{ijk} = \frac{i}{2 \eta^2} \eta \mathring\Gamma_{ijk} \eta
	+ \frac{i}{2 \bar \eta^2} \bar\eta \mathring\Gamma_{ijk} \bar\eta~.
\end{align}
This can be checked by verifying that the $y^i$-variation of \eqref{E:11DefF} matches the $z^i$-variation in \eqref{E:deltaFS.F} using \eqref{E:y.From.z}. Further evidence is provided by checking that the $G_2$ relation \eqref{E:metric} holds upon inserting \eqref{E:11DefF} and the expression for $g_{ij}$ in terms of $\eta$ and $\mathring{g}_{i j}$ from \eqref{E:11DefMetric}.

The holomorphic structure of $F_{ijk}$ suggests that one should identify the bottom component of the chiral superfield $\Phi_{ijk}$ as
\begin{align}
\Phi_{ijk} = \mathring C_{ijk} - \frac{1}{\bar \eta^2} \bar\eta \mathring \Gamma_{ijk} \bar\eta~,
\end{align}
where $\mathring C_{ijk}$ is the 3-form descending from 11D supergravity. Matching transformation rules confirms this, which means that the 3-form $C_{ijk}$ defined by the real part of $\Phi_{ijk}\vert$ actually differs from $\mathring C_{ijk}$, just as $g_{i j}$ defined from $F_{ijk}$ differs from $\mathring g_{i j}$. In both cases, adopting the gauge where $\eta$ is real (equivalently, where $H_i\vert$ vanishes) they become equal.

Let us end on one particularly interesting result that we have not completely understood. The Hermitian metric $\cG_{i j}$ appearing in the gravitino transformation possesses an inverse
\begin{align}
(\cG^{-1})^{ij}
	&= \hat \cF^{-2} g^{ij} - \frac{1}{4} \hat \cF^2 \hH^i \hH^j
		- \frac{i}{2} F^{ijk} \hH_{l}~.
\end{align}
Remarkably, it is this inverse, rather than $\cG_{i j}$ itself, which has an elegant interpretation in 11D. We find simply
\begin{align}
(\cG^{-1})^{ij}
	&= \hat \cF^{-1} \Big(
	\mathring{g}^{i j} + \eta \mathring{\Gamma}^{i j} \bar \eta
	\Big)~.
\end{align}

\section{The conformal supergravity supercurrent}
\label{E:CSGcoupling}
Until this point, we have been treating 4D $N=1$ conformal supergravity as strictly $y$-independent. We have also implicitly assumed that the super-vielbein was invariant under the extended supersymmetry transformations to the order we are working. That is, we have assumed $\delta_1 E_M{}^A = 0$ (equivalently, $\delta_1 H_{\alpha \dalpha} = 0$). This is to be expected, since the component vielbein always transforms with second-order gravitino weight, that is, $\delta e \sim \eps \psi$ into the component gravitino $\psi$ with SUSY parameter $\eps$. This means we have not yet actually determined that the first-order gravitino coupling is consistent with supergravity at the non-linear level.

To remedy this, we will introduce the prepotential superfield $H_{\alpha\dalpha}$ to describe $y$-dependent fluctuations around the $y$-independent background vielbein $E_M{}^A$. The schema for introducing prepotentials to deform a background (non-flat) geometry can be found in refs. \cite{Gates:1983nr, Buchbinder:1998qv}. (See \cite{Butter:2009wy} for the particular case of $N=1$ conformal superspace.) The prepotential $H_{\alpha \dalpha}$ is now subject to the gauge transformations
\begin{subequations}\label{E:deltaL}
\begin{align}\label{E:deltaL.Ha}
\delta H_{\alpha \dalpha}  = \bar \nabla_{\dalpha} L_\alpha - \nabla_\alpha \bar L_{\dalpha} ~,
\end{align}
where the superspace derivative $\nabla_\alpha$ is defined in the $y$-independent background $E_M{}^A$. One must assign $L_\alpha$ transformations to the other potentials. The right choices can be determined following for example \cite{Butter:2009wy}, and correspond to the (covariantized) transformations
\begin{align}
\Delta_0 \Phi &= -\frac{i}{2} \bar\nabla^2 (L^\alpha \nabla_\alpha F )~, \\
\Delta_0 V &= -L^\alpha W_\alpha - \bar L_\dalpha \bar W^\dalpha
	+ i L^\alpha \imath_{\cW_\alpha} F - i \bar L_\dalpha \imath_{\bar\cW^\dalpha} F~, \\
\Delta_0 \Sigma_\alpha &= \frac{i}{2} \bar\nabla^2 (L_\alpha H)~, \\
\Delta_0 X &= \nabla^\alpha (L_\alpha G) + \bar \nabla_\dalpha (\bar L^\dalpha \bar G)
	+ i L^\alpha \imath_{\cW_\alpha} H - i \bar L_\dalpha \imath_{\bar\cW^\dalpha} H~, \\
\delta_0 \cV &= -L^\alpha \cW_\alpha - \bar L_\dalpha \bar \cW^\dalpha~.
\end{align}
In addition, as in the linearized case \cite{Linch:2002wg, Becker:2017zwe}, we must assign an $L_\alpha$ transformation to the gravitino,
\begin{align}
\label{E:deltaL.Psi}
\delta_{-1}\Psi_\alpha = 2 i\, \partial L_\alpha~.
\end{align}
\end{subequations}
We have labeled these by gravitino weight. 
In principle, each of the equations \eqref{E:deltaL} may possess higher-order gravitino modifications on the right-hand sides. 
We should also mention here that $L_\alpha$ itself possesses a certain gauge-for-gauge symmetry where it can be shifted by a chiral spinor superfield; this shift is countered in \eqref{E:deltaL.Psi}, for example, by a shift in $\Xi_\alpha$.

The first order coupling of $H_{\alpha\dalpha}$ to the non-linear action is
\begin{align}
S_{H_a} = \frac{1}{\kappa^2} \int d^4 x \int d^7y \int d^4\theta \,E  \,
	H^{\dalpha \alpha} J_{\alpha\dalpha}
\end{align}
where $J_{\alpha\dalpha}$ can be interpreted as the supercurrent. 
We derive it directly by requiring gauge invariance to lowest order in the $L_\alpha$ transformations. 
This is a long calculation that can be split into two parts. 
The first arises from the minimal coupling of $H_{\alpha\dalpha}$ to the Chern-Simons action and is
naturally written as a 6-form,
\begin{align}
J^{\rm{CS}}_{\alpha \dalpha} &= 
	\frac{i}{4} \nabla_\alpha F \wedge \bar\nabla_\dalpha F \wedge H
	- \frac{1}{2} W_\alpha \wedge \bar W_\dalpha \wedge F
	\eol & \quad
	- \frac{i}{2} W_\alpha \wedge \imath_{\bar \cW_\dalpha} F \wedge F
	- \frac{i}{2} \bar W_\dalpha \wedge \imath_{\cW_\alpha} F \wedge F
	\eol & \quad
	- \frac{1}{3} \imath_{\cW_\alpha} F \wedge \imath_{\bar\cW_\dalpha} F \wedge F~,
\end{align}
or equivalently as a density
\begin{align}
J^{\tiny\rm{CS}}_{\alpha \dalpha} &= 
	-\frac18 \sqrt{g}\, W_{\alpha i j} \bar W_{\dalpha kl} \tilde F^{ijkl}
	+\frac i2 \sqrt{g}\, W_{\alpha i j} \bar\cW_\dalpha{}^k F_k{}^{i j}
	+\frac i2 \sqrt{g}\, \bar W_{\dalpha i j} \cW_\alpha{}^k F_k{}^{i j}
	+2 \sqrt{g}\, \cW_{\alpha}{}^i \bar \cW_\dalpha{}^j g_{i j}
	\eol & \quad
	+ \frac{i}{4 \cdot 3! \cdot3!} \epsilon^{ijklmnp} \nabla_{\alpha} F_{ijk} \bar\nabla_\dalpha F_{lmn} H_p~.
\end{align}
The contribution from the K\"ahler term is more complicated and given by
\begin{align}
J^{\rm K}_{\alpha \dalpha}
	&=
	\frac{1}{2} [\nabla_\alpha, \bar\nabla_\dalpha] \Big(( G \bar G)^{1/3} \sqrt g (\hat \cF - \tfrac{3}{2} \cF)\Big)
	- \frac{3}{2} (G \bar G)^{-1/3} \sqrt g\, \cF' H^i [\nabla_\alpha, \bar\nabla_\dalpha] H_i
	\eol & \quad
	+ \frac{3}{4} \,
		\frac{\pa}{\pa F_{ijk}} (\sqrt g \, \cF (G \bar G)^{1/3})\,
		[\nabla_\alpha, \bar\nabla_\dalpha] F_{ijk}
	\eol & \quad
	+ \frac{i}{2} (G \bar G)^{1/3} \sqrt g\, \hat\cF\, \nabla_{\alpha\dalpha} \log (G / \bar G)
	- \bigg( 3 \bar\nabla_\dalpha \Big(
	(G \bar G)^{-1/3} \sqrt g\, \cF' H^i \nabla_\alpha H_i
	\Big) + \HC \bigg)
	\eol & \quad
	- \cF' (G \bar G)^{-1/3} \Big(
	3 H^i \bar \cW_\dalpha{}^j W_{\alpha i j}
	-3 H^i \cW_\alpha{}^j \bar W_{\dalpha i j}
	- 6 i H^i \cW_\alpha{}^j \bar\cW_\dalpha{}^k F_{i j k}
	\Big)~.
\end{align}

The two quantities $J_{\alpha\dalpha} = J^{CS}_{\alpha\dalpha} + J^K_{\alpha\dalpha}$,
given above, and $J_{\alpha}{}^i$ in \eqref{E:DefJ.v1} describe the two supercurrents
of 11D supergravity written in $N=1$ language. $J_{\alpha\dalpha}$ is the $N=1$
conformal supergravity supercurrent, and defines the first-order coupling to the
$y$-dependent fluctuation superfield $H_{\alpha\dalpha}$. Similarly, $J_\alpha{}^i$
is the extended gravitino supercurrent, describing the first-order coupling to
$\Psi_{\alpha i}$. When the covariant $N=1$ superfields obey their equations of motion,
these currents are subject to the conservation conditions
\begin{subequations}
\begin{align}
\bar \nabla^2 J_{\alpha}{}^i &= 0~, \\
\bar G \nabla^\alpha (\bar\cG_{i j} J_\alpha{}^j)
	&= - i \bar W_{\dalpha\, i j} \bar J^{\dalpha j}~,\\
\bar\nabla^\dalpha J_{\alpha \dalpha} &=  2 i \,\pa_i J_\alpha{}^i
\end{align}
\end{subequations}
These conservation equations are a direct consequence of the gauge transformations
\eqref{E:DeltaGrino}, \eqref{E:deltaL.Ha}, and \eqref{E:deltaL.Psi}.

\section{Toward higher-order terms and a more covariant formulation}
\label{S:CovariantForm}

The ability to couple $\Psi_{\alpha i}$ and $H_{\alpha\dalpha}$ to the non-linear action at first order provides a strong check of consistency. Their couplings correspond to the $N=1$ supercurrent and extended supersymmetry supercurrents of 11D supergravity. The associated lowest-order extended SUSY transformations completely determine the function $\cF$ in the K\"ahler action and lead to a number of consistency conditions. In this section, we will describe how these couplings could be taken to all orders.

One key feature that was useful in determining the gravitino supercurrent $J_\alpha{}^i$ was the simplicity of the $\Xi$ part of transformations of both the gravitino \eqref{E:DeltaGrino} and the other prepotentials \eqref{E:DeltaPot}. These are differential form transformations involving neither the metric $g_{ij}$ nor the function $\cF$ appearing in the K\"ahler action, and so the K\"ahler and Chern-Simons actions must be canceled separately. 
If this feature holds to all orders, it would mean that in order for the K\"ahler action to be canceled by gravitino terms, it must be possible to construct new field strengths $\bm F_{ijk}$, $\bm H_i$, and $\bm G$ by introducing $\Psi$-modifications of the old field strengths so that $\Xi$ invariance is manifest.

To lowest order, this is precisely how the K\"ahler part of the supercurrent \eqref{E:DefJ.v1} arises.
Indeed, the supercurrent \eqref{E:DefJ.v1} can be rewritten as
\begin{align}
\label{E:DefJ.v2}
J_{\alpha}{}^i
	&=
	\Big(\frac{3i}{2} W_{\alpha j k} \frac{\pa}{\pa F_{ijk}}
	+ \cW_\alpha{}^i \frac{\pa}{\pa \log(G \bar G)} 
	+ \frac{i}{2} G \nabla_\alpha \frac{\pa}{\pa H_i} \Big)\Big(- 3\sqrt g (G \bar G)^{1/3} \cF \Big)
 	\cr & \quad
	+ \frac{1}{144} \epsilon^{ijklmnp} F_{jkl} \Big(G \,\nabla_\alpha F_{mnp}
	- 3 \,H_m W_{\alpha np} \Big) 
~.	
\end{align}
Thus, one simply makes the following shifts in $S_K$,
\begin{subequations}
\label{E:CovFSs}
\begin{align}
G \rightarrow \bm G &= G+G \imath_{\cW^\alpha} \Psi_\alpha ~, \\
H \rightarrow \bm H &= H 
	-\frac{i}{2} \nabla^\alpha (G \Psi_\alpha)
	+\frac{i}{2} \bar\nabla_\dalpha (\bar G \bar \Psi^\dalpha)~, \\
F \rightarrow \bm F &= F
	+ \frac{i}{2} \Psi^\alpha \wedge W_\alpha
	- \frac{i}{2} \bar \Psi_\dalpha \wedge \bar W^\dalpha~,
\end{align}
\end{subequations}
which are $\Xi$-invariant to lowest order, and expands to first order in $\Psi$.

This observation cannot be extended simply by exponentiation.
Remarkably, however, it is possible to construct higher-order modifications that ensure
$\Xi$ invariance. For example,
\begin{subequations}
\begin{align}
\bm G - G &= G \imath_{\cW^\alpha} \Psi_\alpha + \cO(\Psi^3)~, \\
\bm H - H &= -\frac{i}{2} \nabla^\alpha (G \Psi_\alpha)
	+\frac{i}{4} G\,\imath_{\bar\cW_\dalpha} (\bar\nabla^\dalpha \Psi^\beta \wedge \Psi_\beta)
	+ \HC
	+ \cO(\Psi^3)~, \\
\bm F - F &=
	+ \frac{i}{2} \Psi^\alpha \wedge W_\alpha
	+ \frac{i}{16} \bar\nabla^2 \Psi^\alpha \wedge \Psi_\alpha \wedge H
	- \frac{i}{8} \bar\nabla^\dalpha \Psi^\alpha \wedge \Psi_\alpha \wedge \bar\nabla_\dalpha H
	\\
	& \quad
	+ \frac{i}{4} \Psi^\alpha \wedge \pa \Psi_\alpha\, G 
	- \frac{i}{4} \Psi^\alpha \wedge \Psi^\beta \wedge \imath_{\cW_\alpha} W_{\beta}
	- \frac{i}{4} \Psi^\beta \wedge \imath_{\cW^\alpha} \Psi_\beta  \wedge W_\alpha
	+ \HC
	+ \cO(\Psi^3)~,
\nonumber
\end{align}
\end{subequations}
correspond to modified $\Xi$-invariant field strengths, provided we introduce higher-order $\Xi$ transformations of the form\footnote{There is some ambiguity in these transformations corresponding to
the ability to make $\cO(\Psi^2)$ field redefinitions of $\Psi_{\alpha}$.}
\begin{subequations}
\begin{align}
\delta_{1} \Psi_{\alpha}
	&=
	\Psi_{\alpha} \imath_{\cW^\beta} \Xi_\beta
	- \frac{1}{2} \imath_{\cW_\alpha} (\Psi^\beta \Xi_\beta)~,\\
\delta_{2} V &=
	\frac{i}{4} \Psi^\alpha \wedge \Xi_\alpha\, G 
	- \frac{i}{4} \bar\Psi_\dalpha \wedge \bar \Xi^\dalpha\, \bar G ~,\\
\delta_{2} \Phi &=
	\frac{1}{8} \bar\nabla^2 (\Psi^\alpha \wedge \Xi_\alpha \wedge H)~. 
\end{align}
\end{subequations}
Proceeding in this way, one determines the higher-order $\Xi$ transformations order by order.

To apply this logic to the Chern-Simons action, recall that in the component 11D theory one can write the integral $\int C \wedge F \wedge F$ over 11D spacetime $M$ as the integral $\int F \wedge F \wedge F$ over some auxiliary 12-manifold whose boundary is $M$ \cite{Witten:1996md}. 
This can be extended to $N=1$ superspace by taking $Y$ to be the boundary of an 8-manifold $Z$ and integrating a super $[4,8]$-form on $\bm X \times Z$. 
The requisite 12-form was computed in eq.\ (5.49d) of ref. \cite{Becker:2017njd}.
Being the superspace version of $F^3$, it involves only field strength superfields so that we can proceed to apply the procedure above.

Carrying out this program generates a large class $\Psi$ corrections necessary for $\Xi$-invariance. 
However, at the linearized level \cite{Becker:2017zwe}, there are quadratic terms involving $\bar \nabla_\dalpha \Psi_\alpha$ that are already $\Xi$-invariant, and we should expect corrections to these terms in the non-linear theory.
In principle, they can be determined by requiring $L_\alpha$ and/or $\Omega$ invariance.

\section{Outlook}

In this paper we have given the construction of eleven-dimensional supergravity in 4D $N=1$ curved superspace to first non-trivial order in the fields with 4D spin $\geq\frac32$ and to all orders in the remaining fields.
More precisely, we formulated a gravitino superfield expansion of eleven-dimensional supergravity and solved it to leading and next-to-leading order for the action and gauge transformations. 
The consistency of the construction relies on a powerful set of local superconformal symmetries arising from the foliation of the spacetime by $N=1$ superspaces. 
This formulation is well-suited to backgrounds in which the spin $\geq\frac32$ components are $Y$-independent but otherwise arbitrary (assuming vanishing vacuum values for the gravitini) such as warped compactifications with fluxes.

The new local symmetry can be understood by comparing it to eleven-dimensional superspace. 
Reduction of the 32 supersymmetries to 4 is parameterized by a complex spinor. 
Dirac bilinears in this spinor define the expected $G_2$ structure on $Y$, but additionally we find deformed chiral and linear superfields corresponding to conformal, $U(1)$, and special conformal compensators. 
In the superfield description, these compensators are mixed up in multiplets containing physical fields.
Because of this, interactions of the latter can be determined exactly by requiring invariance under the gauge transformations of the former. 

On the other hand, the gravitino superfields share compensating St\"uckelberg-like superfield transformations with some of the physical fields for which the action is determined exactly.
Covariantizing under this part then introduces gravitino corrections to the field strengths appearing in there, which we constructed explicitly at next-to-next-to-leading order. 
What is more, the four-dimensional part of the superconformal symmetry appears in the gravitino transformations through its $Y$ dependence, again in a St\"uckelberg-like shift that can only be canceled by similar logic.
This time it involves the 4D $N=1$ conformal supergravity prepotential $H^a$. 

As a first step in this direction, we have explicitly constructed the 4D $N=1$ supercurrent responsible for this mechanism at lowest order. 
However, working order-by-order in a this prepotential is vastly more complicated, since it couples to everything. 
Instead, one might attempt a more covariant approach in which this field does not appear explicitly. 
This would be in close analogy to how we have treated the Kaluza-Klein gauge field $\mathcal V$ covariantly, hiding it in the superspace connection and manipulating only its field strength $\mathcal W$ explicitly. 
In such an approach, the new covariant object $X_i{}^a$ replacing $\partial_i H^a$ and analogous to the aforementioned field strengths would appear as the curvature components in the superspace commutator $[\nabla_A, \nabla_i]$ where $\nabla_A$ is the conformal superspace connection \cite{Butter:2009cp}.
This supergeometric approach is currently under investigation. 

\section*{Acknowledgements}
We thank Sunny Guha and Daniel Robbins for discussions and collaboration during the early stages of this project. 
This work is partially supported by NSF under grants PHY-1521099 and PHY-1620742 and the Mitchell Institute for Fundamental Physics and Astronomy at Texas A\&M University.
We also thank the Simons Center for Geometry and Physics and the organizers of the September 2017 Workshop on Special Holonomy, where results from this work were reported.

\appendix
\section{4D Superspace and supergeometry} \label{A:Geometry}
Our conventions for 4D $N=1$ superspace follow \cite{Butter:2009cp}, where $N=1$ conformal superspace
was introduced for describing conformal supergravity. The 4D supermanifold is described
by local coordinates $z^M = (x^m,  \theta^\mu, \bar \theta_{\dot \mu})$ and is equipped with
a set of 1-form connections that gauge the $N=1$ superconformal algebra.
These are the super-vielbein
$E_M{}^A$, a spin connection $\Omega_M{}^{ab}$, a $U(1)_R$ connection $A_M$, a dilatation connection
$B_M$, $S$-supersymmetry connections $F_M{}^\alpha$ and $F_M{}_\dalpha$, and
special conformal connections $F_M{}^a$. The covariant derivative $\nabla_A$ is given by
\begin{align}
\nabla_A
	= E_A{}^M \Big(
	\pa_M
	- \frac{1}{2} \Omega_M{}^{a b} M_{b a}
	- B_M \mathbb D
	- A_M \mathbb A
	- F_M{}^A K_A
	\Big)
\end{align}
where $M_{ab}$ is the Lorentz generator, $\mathbb D$ is the dilatation generator,
$\mathbb A$ is the $U(1)_R$ generator, and
$K_A = (K_a, S_\alpha, \bar S^\dalpha)$ collectively denotes the three special
(super)-conformal connections. The algebra of these generators with each other
and $\nabla_A$ can be found in \cite{Butter:2009cp} and matches the global $N=1$
superconformal algebra with $\nabla_A$ identified as the super-translation generator
$P_A = (P_a, Q_\alpha, \bar Q^\dalpha)$. The presence of a non-vanishing super-Weyl
tensor $W_{\alpha\beta\gamma}$ deforms the algebra by introducing curvatures
in the (graded) commutators $[\nabla_A, \nabla_B\}$.
While the lowest anti-commutators are unchanged
\begin{align}
\{\nabla_\alpha, \bar \nabla_\dalpha \} = -2i (\sigma^a)_{\alpha \dalpha} \nabla_a~, \qquad
\{\nabla_\alpha, \nabla_\beta\} = 0~, \qquad
\{\nabla_\dalpha, \nabla_\dbeta\} = 0~,
\end{align}
a dimension-$3/2$ curvature operator is introduced \`{a} la super-Yang-Mills,
\begin{align}
[\nabla_\alpha, \nabla_{\beta \dbeta}]
	= 2 \eps_{\alpha \beta} \bar\cW_{\dbeta}~, \qquad
[\bar\nabla_\dalpha, \nabla_{\beta \dbeta}]
	= 2 \eps_{\dalpha \dbeta} \cW_{\beta}~,
\end{align}
where
\begin{align}
\cW_\alpha
	= - i \,W_{\alpha \beta \gamma} (\sigma^{ab})^{\beta\gamma} M_{ba}
	+ \frac{i}{2} \nabla^\gamma W_{\alpha \beta} S^\beta
	+ \frac{i}{2} \nabla_\dbeta{}^\gamma W_{\gamma \alpha \beta} (\bar\sigma^b)^{\dbeta \beta} K_b~.
\end{align}
The vector-vector curvature $[\nabla_a, \nabla_b]$ can be
found in \cite{Butter:2009cp}.
The superfield $W_{\alpha\beta\gamma}$ is a conformal primary (annihilated by $K_A$)
and chiral (annihilated by $\bar\nabla_\dalpha$) and contains the curvature tensors
of conformal supergravity.

As discussed in \cite{Butter:2009cp} (see also \cite{Gates:1983nr} and
\cite{Buchbinder:1998qv} for the conventional formulations in $N=1$ superspace),
an invariant full superspace integral is built out of a scalar function $\mathscr{L}$ via
\begin{align}
\int d^4x\, d^4\theta\, E\, \mathscr{L}~,
\end{align}
where $E = \text{sdet}{(E_M{}^A)}$ is the superdeterminant (or Berezinian) of the
super-vielbein. Super-diffeomorphism invariance in superspace guarantees supersymmetry
in components. Chiral superspace integrals are built out of chiral superfields
$\mathscr{L}_c$ via
\begin{align}
\int d^4x\, d^2\theta\, \cE\, \mathscr{L}_c~,
\end{align}
where $\cE$ is the chiral superspace measure,
see \cite{Butter:2009cp} for its definition in superspace.
In both cases, the functions $\mathscr{L}$ and $\mathscr{L}_c$ must be
conformal primaries and they must possess appropriate Weyl and $U(1)_R$ weights;
that is, $\mathscr{L}$ must have Weyl weight 2 and be $U(1)_R$ neutral, while
$\mathscr{L}_c$ must have Weyl weight 3 and $U(1)_R$ weight 2.

Full superspace integrals are related to chiral superspace integrals via
\begin{align}
\int d^4x\, d^4\theta\, E\, \mathscr{L}
	=
	-\frac{1}{4} \int d^4x\, d^2\theta\, \cE\, \bar\nabla^2\mathscr{L}~.
\end{align}

Because of the presence of the special (super)conformal connections $F_M{}^A$,
the standard rule for a total covariant derivative is slightly modified. Using
\begin{align}
0 = \int d^4x\, d^4\theta\, \partial_M \Big( E\, V^A E_A{}^M\Big) (-)^m
	= \int d^4x\, d^4\theta\, E \Big(
	\nabla_A V^A - F_A{}^B K_B V^A
	\Big) (-)^a~,
\end{align}
for some $V^A$, it follows that
\begin{align}
\int d^4x\, d^4\theta\, E  \, \nabla_A V^A\, (-)^a
	= \int d^4x\, d^4\theta\, E \,
	F_A{}^B K_B V^A (-)^a~.
\end{align}
So if $V^A$ is not a conformal primary, there is a residual connection term
left over. This actually reflects the fact that in these cases, $\nabla_A V^A$ is
not itself a gauge invariant Lagrangian.

\section{Useful variational expressions}
\label{A:VaryExp}

\subsection{The origin of covariantized transformations}\label{A:CovTfor}
The tensor hierarchy that descends from the 3-form of 11D supergravity is gauged by the non-abelian Kaluza-Klein connection. This is most easily described in differential form notation where the 11D exterior derivative decomposes as $d_{11D} \rightarrow D + \pa + \iota_\cF$.
$D$ is the covariant derivative in four dimensions, $D = d - \cL_\cA$, where $\cA$ is the Kaluza-Klein connection, $\pa$ is the internal derivative in seven dimensions, and $\iota_\cF$ is the interior product on an internal form index with the Kaluza-Klein field strength. The 4-form field strength $G = d C$ descends to the set of five field strengths $G_{[0,4]}, \cdots, G_{[4,0]}$ as
\begin{subequations}
\begin{align}
G_{[0,4]} &= \pa C_{[0,3]}~, \\
G_{[1,3]} &= D C_{[0,3]} + \pa C_{[1,2]}~, \\
G_{[2,2]} &= D C_{[1,2]} + \pa C_{[2,1]} + \iota_\cF C_{[0,3]}~, \\
G_{[3,1]} &= D C_{[2,1]} + \pa C_{[3,0]} + \iota_\cF C_{[1,2]}~, \\
G_{[4,0]} &= D C_{[3,0]} + \iota_\cF C_{[2,1]}~.
\end{align}
\end{subequations}
Arbitrary variations of these field strengths involve varying both $C_{[p,3-p]}$ and the Kaluza-Klein vector, leading to
\begin{subequations}\label{E:DeltaGpq}
\begin{align}
\delta G_{[0,4]} &= \pa \delta C_{[0,3]}~, \\
\delta G_{[1,3]} &= D \delta C_{[0,3]} + \pa \Delta C_{[1,2]}
	- \iota_{\delta \cA} G_{[0,4]}~, \\
\delta G_{[2,2]} &= D \Delta C_{[1,2]} + \pa \Delta C_{[2,1]}
	+ \iota_\cF \delta C_{[0,3]}
	- \iota_{\delta \cA} G_{[1,3]}~, \\
\delta G_{[3,1]} &= D \Delta C_{[2,1]} + \pa \Delta C_{[3,0]}
	+ \iota_\cF \delta C_{[1,2]}
	- \iota_{\delta \cA} G_{[2,2]}~, \\
\delta G_{[4,0]} &= D \Delta C_{[3,0]} 
	+ \iota_\cF \delta C_{[3,1]}
	- \iota_{\delta \cA} G_{[3,1]}~,
\end{align}
\end{subequations}
where
\begin{align}\label{E:DeltaCpq}
\Delta C_{[p,3-p]} := \delta C_{[p,3-p]} + \iota_{\delta \cA} C_{[p-1,4-p]}
\end{align}
are the covariantized transformations of the $p$-forms. The relations \eqref{E:DeltaGpq} can be understood as the general variation of the $G_{[p,4-p]}$ field strengths consistent with the Bianchi identities.

A corresponding set of relations exist for the $p$-form hierarchy written in $N=1$ superspace. There the situation is more subtle because the connections $\cA$ and $C_{[p, 3-p]}$ must be built from prepotential superfields. The superspace analogue of the set of field strength variations \eqref{E:DeltaGpq} is
\begin{subequations}\label{E:DeltaGpqSuper}
\begin{align}
\delta F
	&= \frac{1}{2 i} (\Delta \Phi - \Delta \bar\Phi)
	- \frac{1}{2} \iota_{\delta\cV} (\pa \Phi + \pa \bar\Phi)
	- \pa \Delta V~, \\[1ex]
\delta W_{\alpha}
	&= -\frac{1}{4} \bar\nabla^2 \nabla_\alpha \Big(
	\Delta V
	+ i \,\iota_{\delta\cV} F
	\Big)
	+ \pa \Delta \Sigma_{\alpha}
	+ \frac{i}{2} \bar\nabla^2 \Big(\iota_{\delta\cV} \nabla_\alpha F \Big)
	+ \iota_{\cW_\alpha} \delta \Phi
	- i \cL_{\delta\cV} W_\alpha 
	~, \\[1ex]
\delta H
	&= \frac{1}{2i} (\nabla^\alpha \Delta \Sigma_\alpha - \bar\nabla_\dalpha \Delta \bar\Sigma^\dalpha)
	- \pa \Delta X
	- \omega_{\rm h}(\cW_\alpha, \Delta V)
	\cr  & \quad
	- \nabla^\alpha (\iota_{\delta\cV} W_\alpha - i\, \iota_{\delta\cV} \, \iota_{\cW_\alpha} F)
	- \bar\nabla_\dalpha (\iota_{\delta\cV} \bar W^\dalpha + i\, \iota_{\delta\cV} \,\iota_{\bar\cW^\dalpha} F)
	\cr & \quad
	+ \frac{1}{2} \iota_{\delta\cV} \nabla^\alpha (W_\alpha - 2 i \,\iota_{\cW_\alpha} F)
	+ \frac{1}{2} \iota_{\delta\cV} \bar\nabla_\dalpha (\bar W^\dalpha + 2 i \,\iota_{\bar\cW^\dalpha} F)~, \\[1ex]
\delta G &=
	- \frac{1}{4} \bar\nabla^2 (\Delta X - i \,\iota_{\delta\cV} H)
	+ \iota_{\cW^\alpha} \Delta \Sigma_\alpha
	- i \cL_{\delta \cV} G ~.
\end{align}
\end{subequations}
where the covariantized transformations of the prepotentials was given in \eqref{E:DeltaPot}, analogous to \eqref{E:DeltaCpq}.

\subsection{Arbitrary variations of the Chern-Simons and K\"ahler actions}\label{App:VaryCSK}
The variation of the Chern-Simons action with respect to the tensor hierarchy and Kaluza-Klein prepotentials can be decomposed in terms of a full superspace and chiral superspace piece,
\begin{align}\label{E:deltaSCS}
\kappa^2 \delta S_{CS} &= \int d^4 x \int_Y \int d^4\theta \,E  \,\delta \cL_{CS,D}
	+ \Big(\int d^4 x \int_Y \int d^2\theta \,\cE  \,\delta \cL_{CS,F}
	+ \HC \Big)~, \\[2ex]
\delta\cL_{CS,F} &=
	-\frac{i}{4} \Delta \Phi \wedge \Big(\pa \Phi \,G + \frac{1}{2} W^\alpha \wedge W_\alpha
		- \frac{i}{4}\bar \nabla^2 (F \wedge H) \Big)
	\cr & \quad
	- \frac{i}{4} \Delta \Sigma^\alpha \wedge \Big(
		\pa \Phi \wedge W_\alpha
		- \frac{i}{4} \bar\nabla^2 (F \wedge \nabla_\alpha F)
		\Big)~,\eol[2ex]
\delta \cL_{CS,D} &=
	-\frac{1}{4} 
	\Delta V \wedge \Big(
		\pa \Phi \wedge H
		+ F \wedge \nabla^\alpha W_\alpha
		+ 2 \nabla^\alpha F \wedge (W_\alpha - i \iota_{\cW_\alpha} F)
		+ \HC
	\Big)
	\cr & \quad
	+ \frac{1}{4}\, \Delta X \,(\pa \Phi \wedge F + \HC)
	+ \frac{1}{2} \pa H \wedge \iota_{\delta\cV} F \wedge F
	+ \frac{1}{2} H \wedge \iota_{\delta\cV} \pa F \wedge F
	\cr & \quad
	+ \frac{1}{2} F \wedge \iota_{\delta\cV} \,\iota_{\cW^\alpha} F \wedge \nabla_\alpha F
	+ \frac{1}{2} F \wedge \iota_{\delta\cV} \,\iota_{\bar\cW_\dalpha} F \wedge \bar\nabla^\dalpha F
	+ \frac{1}{6} \iota_{\delta\cV} F \wedge \iota_{\nabla \cW} F \wedge F
	\cr & \quad
	- \frac{i}{2} W^\alpha \wedge F \wedge \iota_{\delta\cV} \nabla_\alpha F
	+ \frac{i}{2} \bar W_\dalpha \wedge F \wedge \iota_{\delta\cV} \bar\nabla^\dalpha F~. \nonumber
\end{align}

For the K\"ahler term, it is easier to give its variation in terms of those of the covariant objects directly,
\begin{align}\label{E:deltaSK}
\kappa^2 \delta S_K &= \int d^4 x \int d^7y \int d^4\theta \,E  \,\delta \cL_{K}~, \\
\delta \cL_{K} &=
	- \frac{1}{144} \epsilon^{i j k l m n p} \,\cU_{ijkl} \delta F_{mnp}
	- (G \bar G)^{1/3} \sqrt{g} (\cF - 2 |H|^2 \cF') \,\delta \log (G \bar G)
	\cr & \quad
	- 6 (G \bar G)^{-1/3} \sqrt{g}\, \cF' \,g^{i j} H_i \delta H_j~,\nonumber
\end{align}
and then employ \eqref{E:DeltaGpqSuper}.


\end{document}